%% file: main.tex
\documentclass[final,5p,times,twocolumn,super]{elsarticle}
\journal{xxx}

\input{head.tex}

\begin{document}

\begin{frontmatter}

\title{\mytitle}

\author[NTNU]{\corina}
\author[NTNU]{\pablo}
\author[CUNI]{\sebas}
\author[CUNI]{\peter}
\author[NTNU]{\rita\corref{cor1}}
\cortext[cor1]{rita.dias@ntnu.no}

\begin{abstract}
Poly(amidoamine) (PAMAM) dendrimers are promising candidates for nucleic acid delivery; however, biocompatibility and transfection efficiency remain a challenge. Here, we investigated how the composition of short peptide tails conjugated to generation 2 PAMAM (G2) dendrimers influence DNA association and condensation across a range of pH values. Using a combination of potentiometric titrations, DNA precipitation assays, and coarse-grained molecular simulations with charge regulation, we show that the ionization of G2 dendrimers is strongly affected by both pH and proximity to DNA. Although charge regulation enhances dendrimer protonation and strengthens DNA association at low pH, DNA condensation by unmodified G2 remains largely insensitive to pH within the studied range.

In contrast, conjugation of a single peptide tail introduces a pronounced pH dependence to DNA condensation. Histidine-containing conjugates exhibit the strongest response, with condensation efficiency decreasing markedly as the pH increases. Simulations reveal that the interaction strength between conjugates and DNA depends on both peptide composition and pH and that histidine-containing peptide tails become nearly neutral at physiological pH, contributing little to DNA binding. While single-conjugate simulations explain the trends in DNA association, they do not fully account for the observed condensation behavior, highlighting the importance of collective effects involving multiple conjugates.

Overall, peptide conjugation transforms G2 PAMAM dendrimers from relatively pH-insensitive DNA-condensing agents into pH-responsive DNA-binding systems. These findings provide molecular-level insight into the interplay between charge regulation, peptide composition, and DNA condensation.

\end{abstract}


\end{frontmatter}

\section{\label{sec:introduction}Introduction}

The therapeutic strategy of modulating genetic expression through the intracellular delivery of polynucleotides  has gained significant importance in modern medicine. A wide range of delivery approaches has been developed to facilitate the transport of nucleic acids into cells. These strategies must address several critical challenges:  protecting nucleic acids from enzymatic degradation,~\cite{Mendes2017DendrimersTherapy} enabling their passage across the cellular membrane and ensuring successful endosomal escape to reach intracellular targets.~\cite{Segura2001MaterialsDelivery} Successful delivery systems must therefore be capable of overcoming these biological barriers to achieve therapeutic efficiency.

Among non-viral delivery systems, cationic lipids and polymers are commonly used due to their ability to complex with and condense nucleic acids. 
Cationic poly(amidoamine) PAMAM dendrimers are an example of such non-viral vectors that have been explored for nucleic acid delivery.\cite{Haensler1993, Bielinska1996RegulationDendrimers, Bielinska1997TheDNA, Waite2009, Santos2010, Tsai2011IntrinsicallyStudy, Yoo2000EnhancedDendrimers} Their water solubility and high charge density enables strong interactions with DNA, leading to DNA condensation and the formation of dendriplexes.~\cite{Bielinska1997TheDNA, Fant2008DNATranscription, Ainalem2009CondensingMorphology, Fu2016DrivingElectrostatic}
Additionaly, the presence of amine groups renders PAMAM dendrimers sensitive to pH variations, which is advantageous for intracellular delivery.
They are also considered biocompatible at low concentrations and low generation.~\cite{Janaszewska2019CytotoxicityDendrimers} To further enhance their biocompatibility, targeting ability, and transfection efficiency, PAMAM dendrimers can also be functionalized with peptides~\cite{ Waite2009,Santos2010, Urbiola2018NovelReceptors,Bae2021ApoptinTherapy, Ebrahimian2022DevelopmentStudies}.
In a previous study, we showed that the net charge of peptides conjugated to generation 2 PAMAM (G2) dendrimers influence their DNA complexation capacity, with fewer conjugates being required to achieve DNA condensation compared to unmodified G2, for positive peptides.~\cite{Dannert2023DNACharge}

Typically, dendriplexes enter cells via endocytosis. As mentioned, one of the main challenges of nucleic acid delivery is the release of DNA from the endosomes into the cytosol, a process known as endosomal escape. 
As the endosomes mature into late endosomes and lysosomes, the pH decreases from approximately 7.4 to 6.2 (late endosome) and eventually to 4.7 (lysosome).~\cite{Scott2014EndosomeFunctions,Pang2020TheNanoparticles} This acidification leads to the increased protonation of weakly basic groups, which is believed to trigger endosomal escape either by increasing osmotic pressure and causing membrane rupture (the proton sponge effect),~\cite{Behr1997TheExploit,Won2009MissingComplexes, Needham1990ElasticCholesterol.,Roy2020LysosomalMicrocapsules} or by destabilizing the endosomal membrane through the formation of pores by highly (positively) charged molecules or complexes.~\cite{Bieber2002IntracellularComplexes,Vaidyanathan2015QuantitativeMembrane,Rehman2013MechanismLysis} 
Both mechanisms facilitate the release of DNA or dendriplexes into the cytosol. 

Therefore, the ability to modulate the charge of the dendrimers and dendriplexes in response to pH changes offers a promising strategy to enhance endosomal escape.

In this work, we investigate how the composition of short peptides conjugated to G2 PAMAM dendrimers influence DNA condensation across varying pH conditions. By combining \textit{in vitro} assays with Langevin dynamics simulations, we show that peptide conjugation amplifies the pH-dependence effects G2 dendrimers on DNA condensation.

\begin{figure*}[htb!]
    \centering
    \includegraphics[width=.9\textwidth]{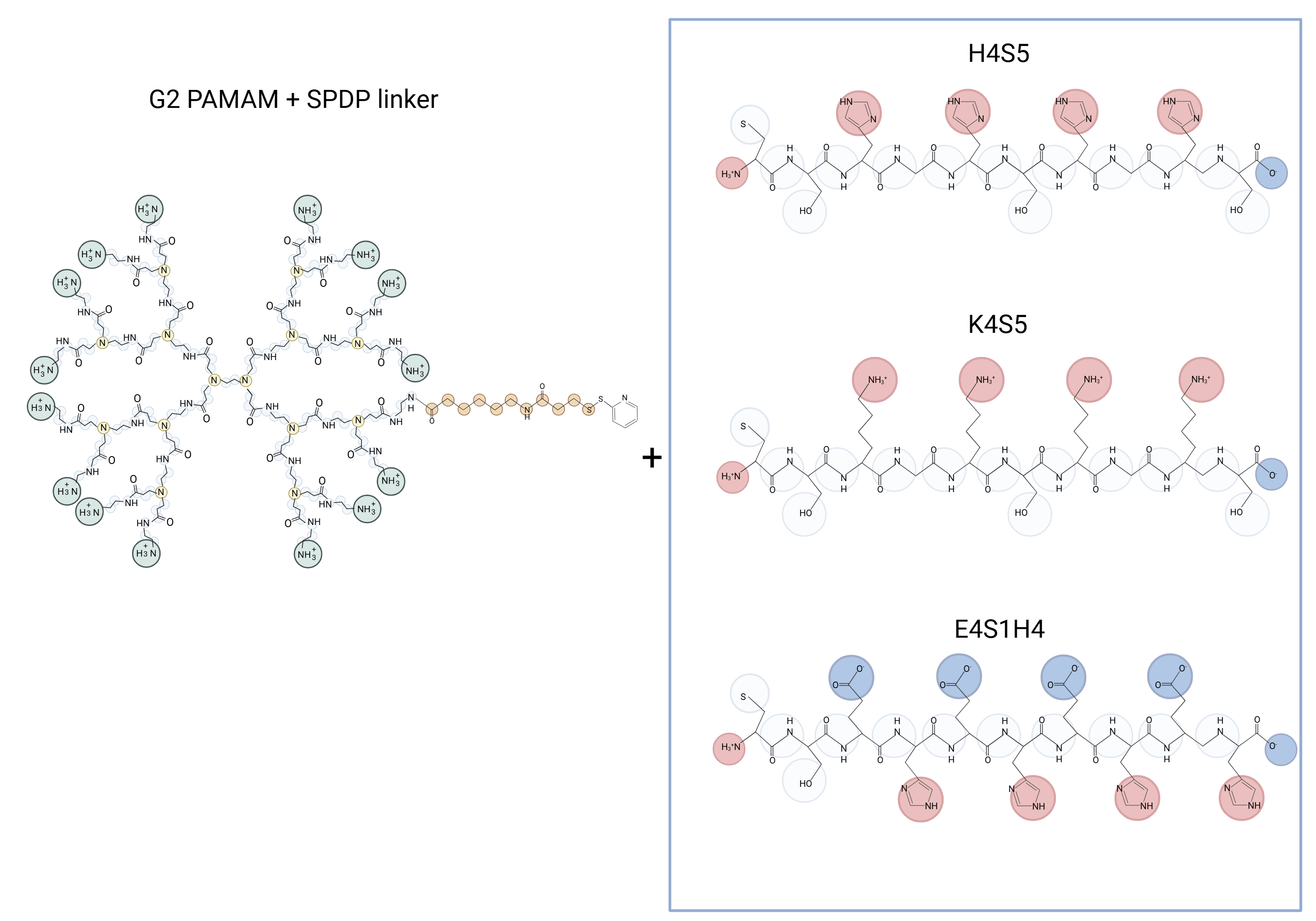}
    \caption{Coarse-grained representation of the G2 PAMAM dendrimer with an attached sulfo-LC-SPDP linker (left) and the peptides used for conjugation (right). Circles represent beads in the coarse-grained model. Green beads indicate the primary amine end groups of the dendrimer, yellow beads represent the internal tertiary amine groups. In the peptides, red beads denote basic residues and blue beads represent acidic residues. Created using BioRender.com.}
    \label{fig:PAMAM_peptides_scheme}
\end{figure*}

\section{Materials and Methods}
\subsection{Materials}
PAMAM dendrimers generation 2.0 with ethylenediamine cores (supplied as 20 wt\% in methanol), phosphate buffered saline (PBS) ta\-blets, dithiothreitol (DTT), sulfo-LC-SPDP, sodium acetate, acetic acid, and Trizma\textregistered base were purchased from Sigma-Aldrich/Merck. Spectrum\textsuperscript{TM} Spectra/\-Por\textsuperscript{TM} Biotech Cellulose Ester dialysis membrane tubing with a MWCO of \SI{0.5}{\kilo Da} was purchased from Thermo Fisher Scientific. Ultrapure water (resistivity \SI{18.2}{\mega\ohm\cm}, Milli-Q plus, Merck Millipore) was employed in the experiments.  

DNA (gwiz-Luc plasmid , 6732bp) was obtained from Aldevron (Fargo, ND).

Custom-designed peptides were purchased from Genscript. The sequences, listed in the right-hand column of Table \ref{tab:tail-charge}, include a cysteine residue to facilitate conjugation. 
Peptides were dissolved in MilliQ water.

To investigate pH-dependent behavior, experiments were conducted using \SI{10}{\milli\molar} sodium acetate buffers at pH 4.3, 5.0 and 5.7, and \SI{10}{\milli\molar} Tris-HCl buffers at pH 7.1 and 8.2. All buffers were adjusted to an ionic strength of \SI{16.9}{\milli\molar}, using NaCl.
\begin{table}[htb]
\centering
\caption{Overview of the peptide amino acid composition used in the experimental work and in the simulations. }

\begin{tabular}{c|c}
    \toprule
    System& peptide a.a. composition \\
    \midrule
    G2-H4S5 &  CSHGHSHGHS\\
    G2-K4S5 &  CSKGKSKGKS \\
    G2-E4S1H4 &  CSEHEHEHEH\\
    \bottomrule
\end{tabular}
\label{tab:tail-charge}
\end{table}

\subsection{PAMAM-Peptide Conjugation}

The conjugation of peptides to G2 PAMAM dendrimers was performed following protocols adap\-ted from Santos et \textit{al.}~\cite{Santos2010} and Waite et \textit{al.}.~\cite{Waite2009}
First, G2 dendrimers were air-dried to remove the methanol and then dissolved in \SI{10}{\milli\molar} PBS buffer. 
Sulfo-LC-SPDP was added to the dendrimer solution at a molar ratio 6:1 (sulfo-LC-SPDP:G2), and the mixture was stirred for 2.5 hours at room temperature to functionalize the dendrimers.
Unreacted SPDP was removed by overnight dialysis against MilliQ water. The samples were then freeze-dried and eluted in \SI{10}{\milli\molar} PBS buffer. 

The degree of SPDP functionalization was assessed using a DTT assay (described below). Peptides were then conjugated to the G2-SPDP complexes by adding them at a 1:1 molar ratio relative to the attached SPDP groups. The reaction was allowed to proceed overnight at room temperature. 

To evaluate conjugation efficiency, DTT assays were performed both before and after peptide addition.
Briefly, \SI{10}{\micro\liter} of a \SI{15}{\milli\gram\per\milli\liter} DTT stock solution in PBS was added to \SI{1}{\milli\liter} of \SI{0.01}{\milli\molar} G2-SPDP or G2-peptide conjugate solution and incubated for 15 minutes at room temperature.
The absorbance at \SI{343}{\nano\meter}, corresponding to pyridine 2-thione (a byproduct of the reaction), was measured using an Agilent 8453 UV/Vis spectrophotometer. This absorbance was used to calculate the average number of SPDP linkers and peptides per dendrimer. On average approximately two SPDP linkers and one peptide were conjugated per dendrimer for all peptide variants. \reffig{fig:DTT assay} in \ref{section:SI_Conjugation_of_PAMAM_G2_and_peptides} shows the pyridine-2-thione absorbance after addition of the linker and after conjugation with the different peptides.

\subsection{Precipitation Assay}
DNA condensation by G2 PAMAM dendrimers and PAMAM-peptide conjugates was assessed via precipitation assays at varying molar ratios (\rmol) and pH values. In each assay, \SI{10}{\micro\liter} of dendrimer or conjugate solutions, at varying concentrations, was added to \SI{4}{\micro\liter} of \SI{0.5}{\gram/\liter} gwiz-Luc DNA in \SI{86}{\micro\liter} of buffer, resulting in a final DNA concentration of \SI{0.02}{\gram/\liter}. The solution was gently pipetted to mix and incubated for \SI{30}{\minute} at room temperature.
Following equilibration, the samples were centrifuged at 3600$g$ for \SI{30}{\minute}. The absorbance of the supernatant at \SI{260}{\nano\meter} was measured using a NanoDrop Spectrophotometer to quantify unbound DNA.

\subsection{Potentiometric Titration}
Prior to titration, G2 dendrimers and G2-peptide conjugates were dialyzed overnight against pure MilliQ water using dialysis tubing with a MWCO of \SI{0.5}{\kilo Da} and freeze-dried. Peptides were delivered freeze-dried by the manufacturer.
Potentiometric titrations of G2, peptides and G2-peptide conjugates were performed in \SI{0.1}{\molar} aqueous HCl by incremental addition of \SI{0.1}{\molar} NaOH until a stable pH value of 12 was reached. A waiting period was included between additions to allow the pH to stabilize.
Titration curves were recorded at \SI{23}{\degreeCelsius} using a Metrohm 888 Titrado Compact titrator equipped with a Pt1000 temperature sensor and a LL Biotrode 3.0 mm glass electrode. 
Before each titration, a blank titration was performed under identical conditions using only \SI{0.1}{\molar} aqueous HCl titrated with \SI{0.1}{\molar} NaOH.

Details on the analysis and processing of the potentiometric titration data are provided in \ref{section:SI_Titration}.

\subsection{Molecular modeling}
To assess how DNA affects the ionization of G2 dendrimers and conjugates, a coarse-grained model that faithfully describes the structure and charge distribution of the double helix was used. First, an all atomistic, helical DNA structure with an arbitrary sequence was generated using UCSF Chimera.~\cite{Pettersen2004UCSFAnalysis} The structure was then coarse grained with the model DNA spanning the entire box length to avoid end effects.
In this representation, each base pair consists of three backbone beads, two for the deoxyribose sugar and one for the phosphate group, which carries a charge of $-1$ and has a radius of $r_{\text{phos}} = \SI{2.9}{\angstrom}$. Purine bases (adenine and guanine) are modeled with four beads, while pyrimidine bases (cytosine and thymine) use three beads, each with a radius of $r_{\text{base}} = \SI{3.35}{\angstrom}$, matching the radius of the deoxyribose beads. All radii are derived from the van der Waals radii of the respective chemical groups. A rigid model was used, fixing the DNA particles in place and thus neglecting DNA bending in the presence of dendrimers. This assumption is justified, as the length of the model DNA in our system is well below the persistence length of DNA, even at high ionic strength.~\cite{Mantelli2011}

The peptides were modeled using a bead-and-spring approach, with two beads per amino acid, one representing the backbone and the other the side chain.~\cite{Lunkad2021QuantitativeOligopeptides} Ionizable side chain beads, along with terminal groups, were assigned p$K_{\text{a}}$ values based on experimentally determined data (see Table \ref{tab:pKa}).

\begin{figure}
    \centering
    \includegraphics[width=\linewidth]{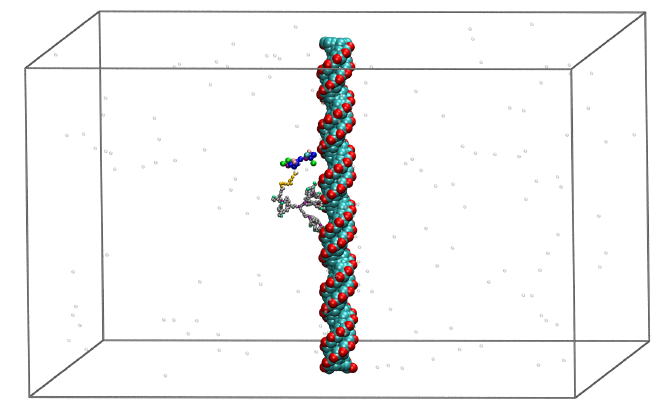}
    \caption{Representative frame from a simulation depicting the G2-H4S5 conjugate in proximity to a fixed DNA double helix. One of the central beads of the dendrimer is fixed at \SI{2.4}{\nano\meter} from the center of the DNA.}
    \label{fig:snapshot_G2-H4S5_DNA}
\end{figure}

In the quasi-atomistic model of the dendrimer, each methyl and amide group was represented by a single, neutrally charged bead. Beads corresponding to tertiary amines (at the branching nodes) and primary amines (at the terminal positions) were modeled with ionizable states that could switch on and off. Their assigned p$K_\text{a}$ values are also listed in Table \ref{tab:pKa}. All dendrimer beads were assigned a radius of $r_{\text{den}}=$\SI{1.5}{\angstrom}.
The linker was modeled similarly, using 11 beads with $r_{\text{link, Me}}=$\SI{1.54}{\angstrom} for the methyl groups and $r_{\text{link, C}}=$\SI{1.5}{\angstrom} for the other groups, as shown in \reffig{fig:PAMAM_peptides_scheme}. Following the chemistry of the reaction, one primary amine of the dendrimer and the sulfur group of the peptide were considered non-ionizable in the conjugates. \reffig{fig:PAMAM_peptides_scheme} illustrates the models of the dendrimer, linker and peptides and \reffig{fig:snapshot_G2-H4S5_DNA} shows a simulation snapshot of the G2-H4S5 conjugate interacting with the fixed DNA double helix.

The systems were studied by combining Langevin Dynamics (LD) simulations to probe the conformational evolution of the dendrimers, peptide, and dendrimer-peptide conjugates and a constant-pH (cpH) Monte Carlo procedure to account for the acid-base equilibrium of the ionizable groups.
Each simulation was run for $7.5\times 10^4$ cycles.
One simulation cycle consisted of 4000 LD integration steps, with a time step $\Delta t=0.005\tau$, followed by approximately 100 cpH titration attempts.
 The first 20\% of each simulation was discarded as equilibration. 
Statistical uncertainty of the computed averages was estimated using the block analysis method.~\cite{Janke2002StatisticalEstimation}

The degree of ionization, $\alpha$, was calculated as the ensemble average over the ionization states throughout the simulation. 
Potential of mean force plots were calculated using a weighted average umbrella sampling.

Full details on the simulation protocol and data processing are provided in the \ref{sec:MolSims}.

The systems and acid-base reactions were set up using the pyMBE molecule builder~\cite{Beyer2024PyMBE:ESPResSo} and the simulations were carried using the ESPReSso v4.2.1 simulation package.~\cite{Weik2019ESPResSoSystems} All computational work was performed on the IDUN computing cluster at NTNU.~\cite{Sjalander2019EPIC:Infrastructure}

\begin{table}[htb]
\centering
\caption{p$K_{\text{a}}$ values used for the ionizable groups in the simulations. The values for the amino acids and peptide termini were obtained from Nozaki et al.,~\cite{Nozaki196784Behavior} while the values for the amine groups of the dendrimer were taken from Cakara et al..~\cite{Cakara2003MicroscopicTitrations}}

\begin{tabular}{c|c}
    \toprule
    Group & p$K_{\text{a}}$ \\
    \midrule
    PAMAM primary amine &  9.15\\
    PAMAM tertiary amine &  6.00 \\
    N-terminus & 7.5\\
    C-terminus & 3.8\\
    Histidine& 6.3\\
    Lysine &  10.4\\
    Glutamic acid & 4.4 \\
    Cysteine & 9.5\\ 
    \bottomrule
\end{tabular}
\label{tab:pKa}
\end{table}

\section{Results and Discussion}

The validity of the ionization behavior of the G2 PAMAM model has been established previously.\cite{PAMAM_Subm} Here the ionization behavior of the peptides is examined, along with the effect of pH on the individual interactions of G2 and peptides with DNA. Next, we extend the study to the titration and DNA binding of PAMAM-peptide conjugates. Lastly, we explore the condensation of DNA mediated by PAMAM-peptide conjugates at varying pH.

\subsection{G2 PAMAM Dendrimers Condense DNA Independently of pH}

\begin{figure}
    \centering
    \includegraphics[width=\linewidth]{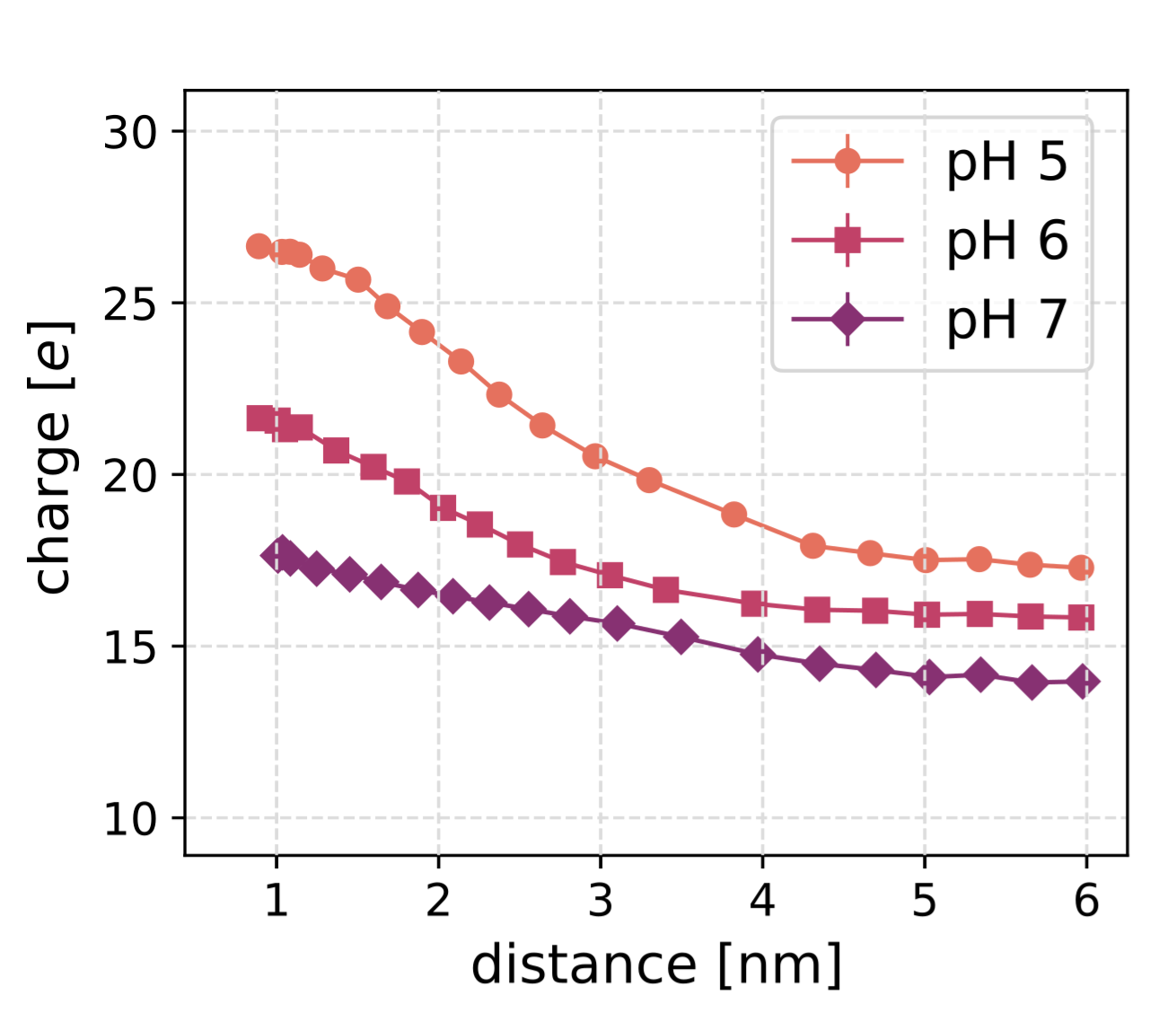}
    \caption{Total charge of dendrimers as a function of distance from the center of the DNA at three different pH values, obtained through molecular modeling.}
    \label{fig:PAMAM_charge_DNA}
\end{figure}

\reffig{fig:PAMAM_charge_DNA} shows the impact of a DNA molecule on the ionization behavior of PAMAM dendrimers, as determined through molecular modeling. We focus on the pH range 5-7, which is particularly relevant for endosomal escape. As expected, the overall charge of the dendrimers decreases with increasing pH. Further, at all pH values studied, the dendrimer charge increases as it approaches the DNA. This effect, known as charge regulation, is most pronounced at pH 5. Similar charge regulations behavior has been previously observed in other polyelectrolyte systems.~\cite{Pineda2024ChargePolyelectrolytes, Landsgesell2019SimulationsGels, Stornes2017MonteConcentration, Ulrich2005ComplexationInfluences, Rathee2018WeakCharging, NarayananNair2017ComplexationPolyampholytes,beyer2025charge}

\begin{figure}
    \centering
    \includegraphics[width=\linewidth]{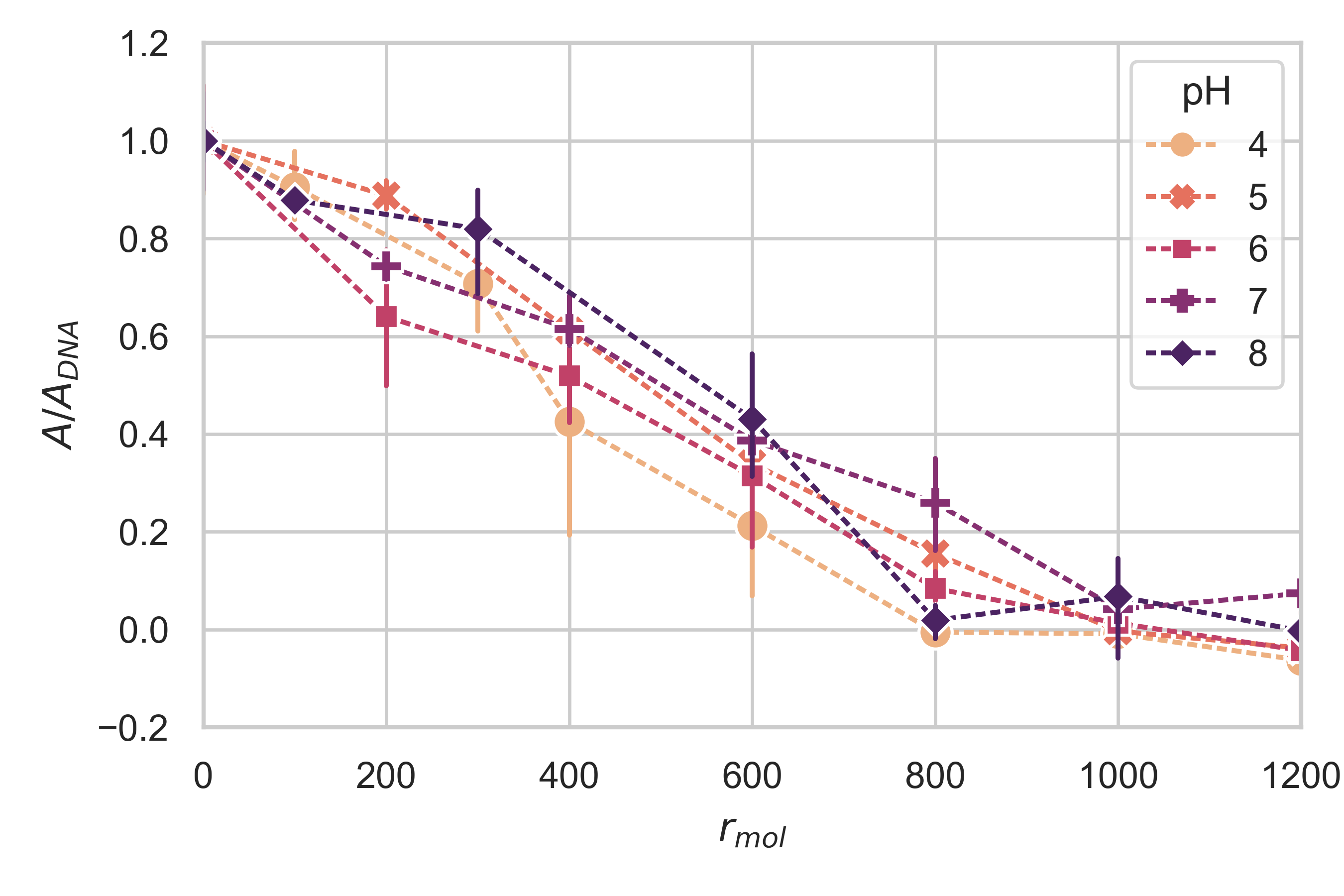}
    \caption{Normalized DNA absorbance from precipitation assay with G2 dendrimers at varying $r_{\text{mol}}$ and pH.}
    \label{fig:precipitation_G2}
\end{figure}

Precipitation assays were performed to evaluate the interaction between G2 PAMAM dendrimers and DNA under varying buffer pH conditions (see \reffig{fig:precipitation_G2}). The results indicate no significant differences in DNA precipitation efficiency across the tested pH range, although a slight trend towards reduced condensation efficiency at higher pH values is observed. Nevertheless, a molar ratio of $r_{\rm mol}>800$ is generally sufficient to induce DNA condensation.
In a previous study, gel electrophoresis showed complex formation between G2 PAMAM dendrimers and DNA at approximately $r_\text{mol} = 400$, while dye exclusion experiments revealed a reduction in normalized fluorescence intensity to 0.55 at $r_\text{mol}=1550$, indicating substantial inhibition of dye binding to DNA.~\cite{Dannert2023DNACharge} Both precipitation assays and gel electrophoresis assess global condensation effects, involving multiple DNA strands, whereas dye exclusion probes more localized interactions between individual dendrimers and DNA. Therefore, the similarity in results between precipitation and gel electrophoresis is expected.  
The apparent pH-independence of DNA condensation may be attributed to the relatively high net charge of the dendrimers, which remains substantial even at high pH. 
As shown in \reffig{fig:PAMAM_charge_DNA}, simulation results reveal that although the total charge of G2 dendrimers decreases with increasing pH, the minimum charge remains at $+14e$, sufficient to drive DNA condensation.

\subsection{Short Peptides are not Sufficiently Charged to Condense DNA}

\begin{figure}
    \centering
    \includegraphics[width=\linewidth]{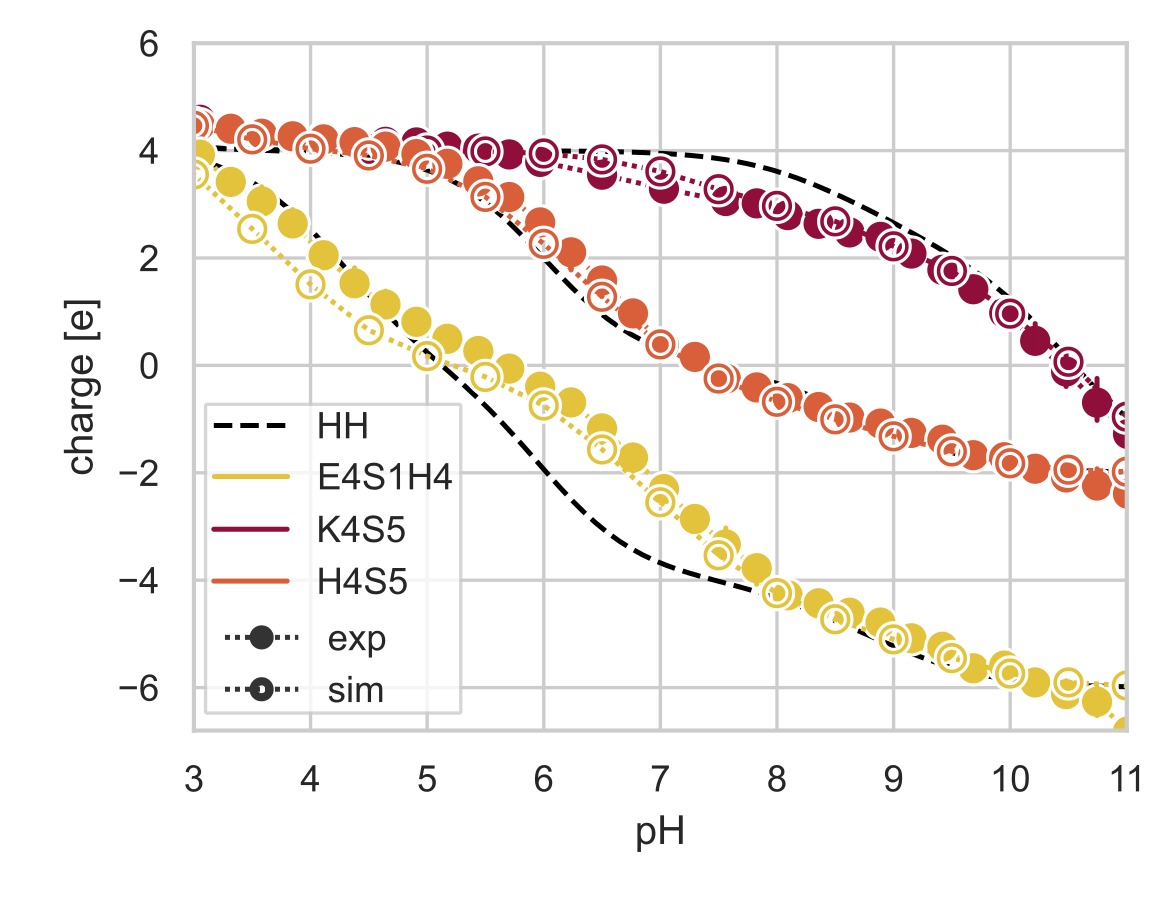}
    \caption{Potentiometric titration of peptides (filled symbols) compared to the ionization of the peptides in the simulations (open symbols). HH refers to the ideal solution using the Hendersson-Hasselbalch equation.}
    \label{fig:titration_peptides}
\end{figure}

Similarly to G2 dendrimers, the ionization of the coarse-grained peptide models, described using a two-bead approach per amino-acids,~\cite{Lunkad2021QuantitativeOligopeptides} is found to be in good agreement with the potentiometric titrations, as shown in \reffig{fig:titration_peptides}. 
The peptide E4S1H4 shows the largest deviation between experiments and simulations, in pH range 3-5. 
At $\text{pH}>10$ the experimental results become less reliable, due to absorption of $\text{CO}_2$, which explains the deviations at high pH values. 

All titration curves show distinct regions of increased ionization from high to low pH, that align well with the known \pKa\ values of the peptide residues. For example, H4S5 (orange circles in  \reffig{fig:titration_peptides}) shows a sharp increase in net charge around pH 6.3, corresponding to the \pKa\ of histidine, followed by a more gradual increase near pH 8.35, which is consistent with the \pKa\ of cysteine.

\reffig{fig:titration_peptides} also shows deviations between the analytical prediction based on the Henderson–Hassel\-balch equation (\refeq{eq:HH} in \ref{section:SI_Titration}) and the simulation and experimental data for peptides K4S5 and E4S1H4. These deviations occur due to the so-called polyelectrolyte effect, where ionization of individual monomers is suppressed due to electrostatic repulsion between the monomers.~\cite{Reed1992MontePolyelectrolytes, Ullner1994ConformationalArguments, Blanco2023UnusualPolyelectrolytes} This is especially evident in the simulation data of the E4S1H4 peptide (open yellow circles in \reffig{fig:titration_peptides}) where the analytical prediction overestimates the (positive) charge at low pH values and overestimates the negative charge between pH 5 and 8, with the data crossing at pH 5, where the peptide is neutral.

The ability of the peptides to condense DNA was evaluated across the pH range 4-8, up to $r_{\rm mol} = 1200$. This corresponds to a charge ratio (N/P) of approximately 4, based on the peptides carrying a net charge of $+4e$, specifically  H4S5 at pH 4 and K4S5 in the pH range 4-6.

Given the relatively low net charge and charge density of these peptides, even at acidic conditions, it is not surprising that no DNA condensation was observed up to $r_{\rm mol} = 1200$ (\reffig{fig:precipitation_K4S5} in \ref{section:Prec_peptides}). Higher concentrations of short polyelectrolytes are typically required to induce DNA condensation.~\cite{ras98:381,Takahashi1997DiscretePolyamines} While the binding of short polyelectrolyte chains to DNA is thermodynamically favorable due to the counterion release, actual DNA condensation, driven by correlation effects and bridging interactions,~\cite{gul84:2221, kha99:121, Khan1999AnomalousTheory} results in a reduction of the DNA's conformational entropy and, to some extent, the translational entropy of the polycations along the DNA backbone.~\cite{dia03:8150}

\subsection{G2 PAMAM-Peptide Conjugates Exhibit pH-Sensitive DNA Association}

\begin{figure}
\centering
\includegraphics[width=\linewidth]{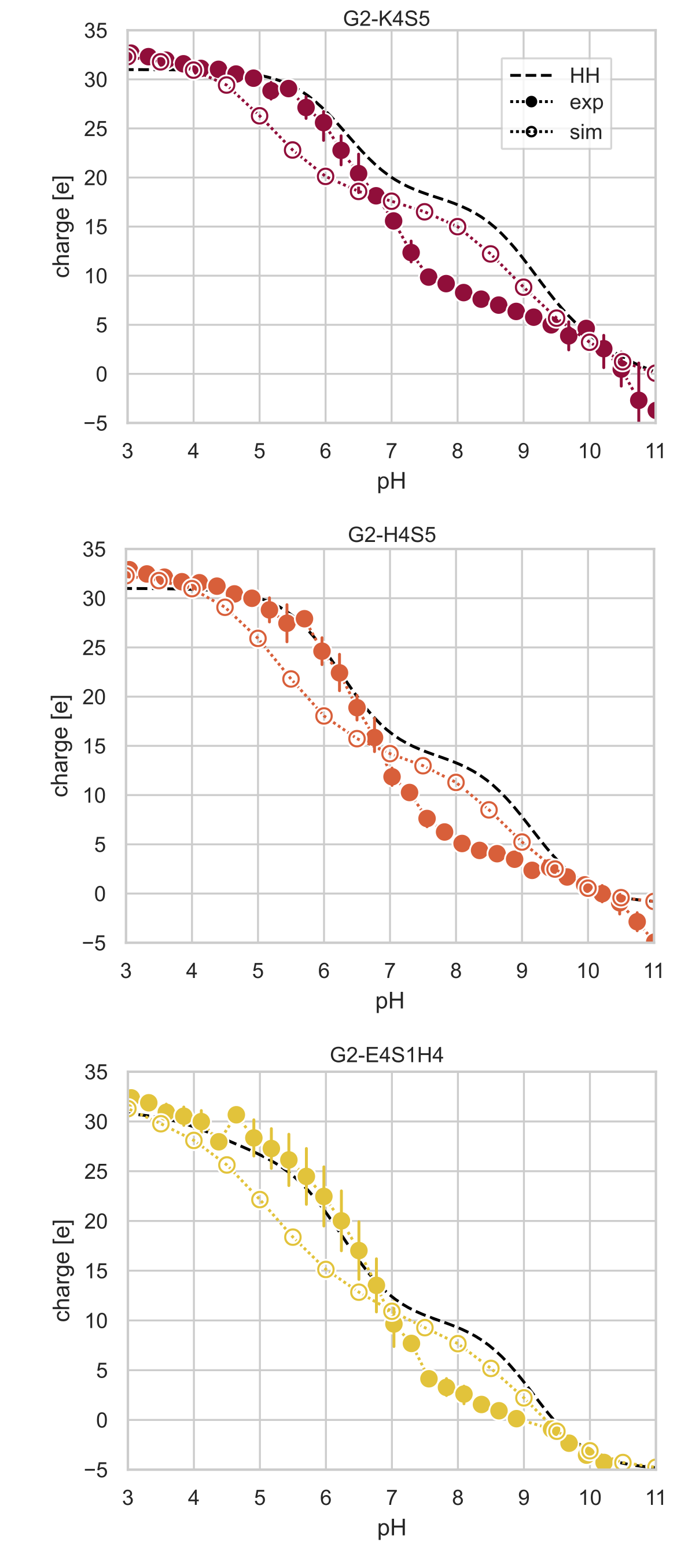}
\caption{Potentiometric titration of conjugates (filled circles) compared to the ionization of the conjugates in the simulations (open circles). HH refers to the ideal solution with the Hendersson-Hasselbalch equation.}
\label{fig:titration_conjugates}
\end{figure}

We now turn to the effect of pH on the interaction between G2 PAMAM-conjugates and DNA, beginning with an analysis of their ionization behavior. In  \reffig{fig:titration_conjugates} we compare the experimental titration data (full circles), simulation results (open circles), and analytical predictions. 
As observed for unmodified G2 PAMAM, the ionization of the amine groups in the conjugates is suppressed across most of the pH range when compared to the ideal behavior predicted by the Henderson-Hasselbalch equation. Interestingly, and in contrast to the behavior of the individual dendrimers and peptides, the potentiometric titration data show poor agreement with the simulation results.

Across all systems studied, the conjugates exhibit near-ideal ionization behavior under acidic conditions, with the simulations underestimating the net charge. However, in the pH range of 7 to 9.5, the experimental data reveal a pronounced suppression of charge relative to both simulations and analytical predictions. 

The deviation observed between the experimental titration curves and simulations results may be attributed to the presence of multivalent phosphate ions originating from the phosphate buffer used during peptide conjugation. 
$^{31}$P NMR analysis confirms that, despite extensive dialysis in pure water, the G2-peptide conjugate samples retained a non-negligible amount of phosphate ions, with P:conjugate molar ratios ranging from \num{12.7} to \num{44.5} depending on the conjugate (see Fig. \ref{fig:PNMR} and Tables \ref{tab:NMR_areas} and \ref{tab:NMR_ratio} in \ref{section:PNMR}). 
Furthermore, experimental studies \cite{keri2017} and molecular modeling \cite{PAMAM_Subm} have shown that phosphate ions accumulate around PAMAM dendrimers. Although the effect of phosphate binding on the ionization state of G2 is expected to be modest, it complicates the interpretation of the potentiometric titration data.

The presence of phosphates is also expected to affect the interactions of the conjugates with DNA. From an electrostatic perspective, the divalent phosphate ions are likely to be displaced from the G2-peptide conjugates upon association with DNA. However, the overall binding affinity may decrease because the pre-associated phosphate ions reduce the counterion-release entropy gain that normally accompanies complex formation. Since all the DNA binding experiments are conducted in a phosphate buffer, the influence of phosphate ions on the interaction with DNA does not contribute to the observed differences between the conjugates.
While phosphate ions are not explicitly included in the simulations, the different observables investigated experimentally and computationally limit the comparison to qualitative trends.

Despite peptide modifications, all conjugates retain  a substantial net charge in the pH range 5-7. At pH 7 it varies from approximately $+15e$ for G2-K4S5 and G2-H4S5 (similar to unmodified G2) to around $+10e$ for G2-E4S1H4. 

\begin{figure}
    \centering    
    \includegraphics[width=\linewidth]{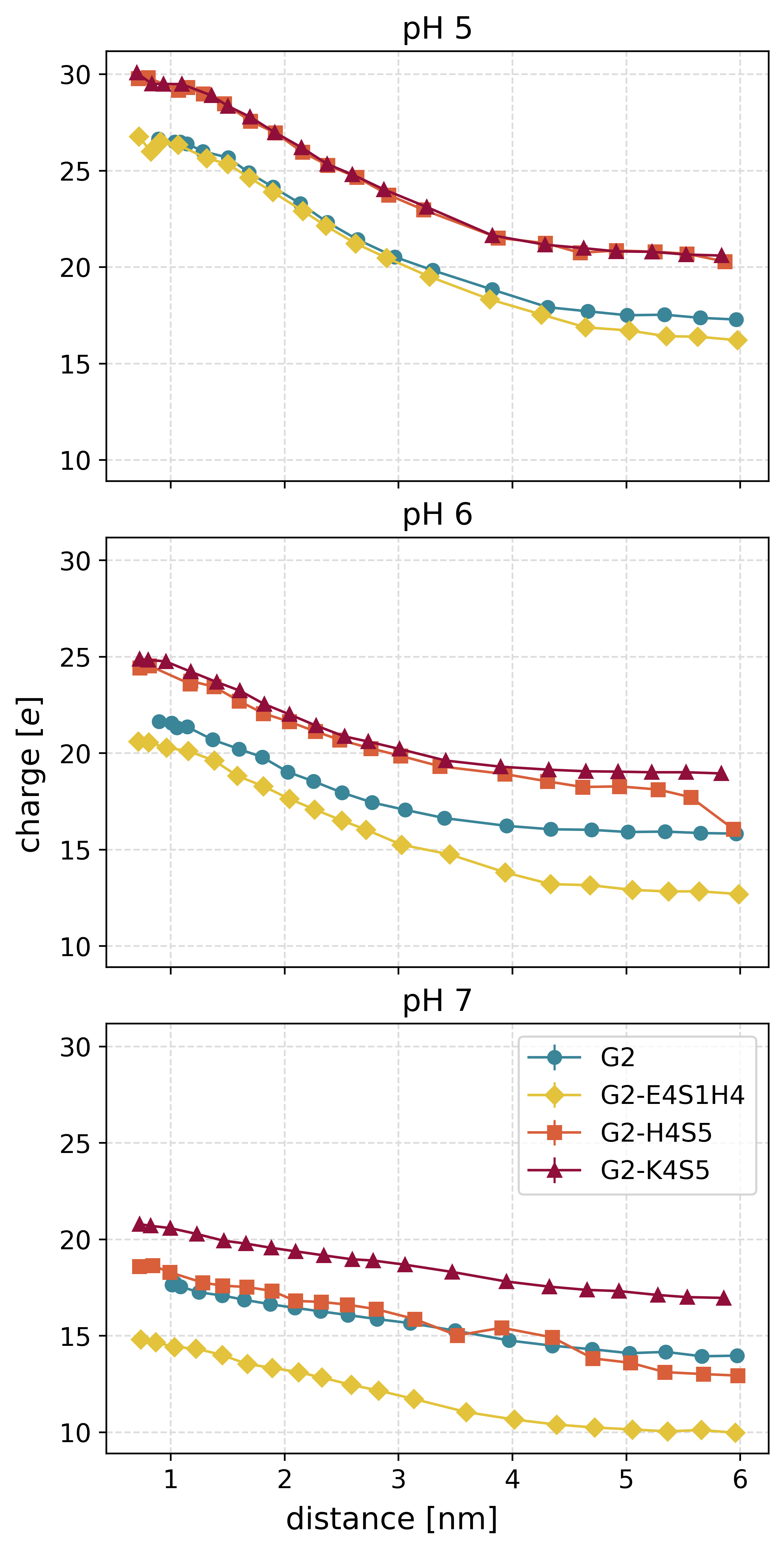}
    \caption{Total charge of dendrimers/conjugates in the simulations as a function of distance from the center of the DNA at three different pH values.}
    \label{fig:charge_pH}
\end{figure}

As shown above for G2 PAMAM, the presence of DNA significantly affects the charge and ionization behavior of all titratable groups in the conjugates. 
The computed net charge of the conjugates as function of distance from the DNA center, across the pH range 5-7, is shown in \reffig{fig:charge_pH}, alongside the corresponding data for G2.
In all systems, the net charge increases as the conjugates approach the DNA. This effect is more pronounced at the lowest pH, where the difference in ionization between DNA-associated and free conjugates/dendrimers reaches approximately $10e$, or about 30\% of the total charge. 

\begin{figure}
    \centering
    \includegraphics[width=\linewidth]{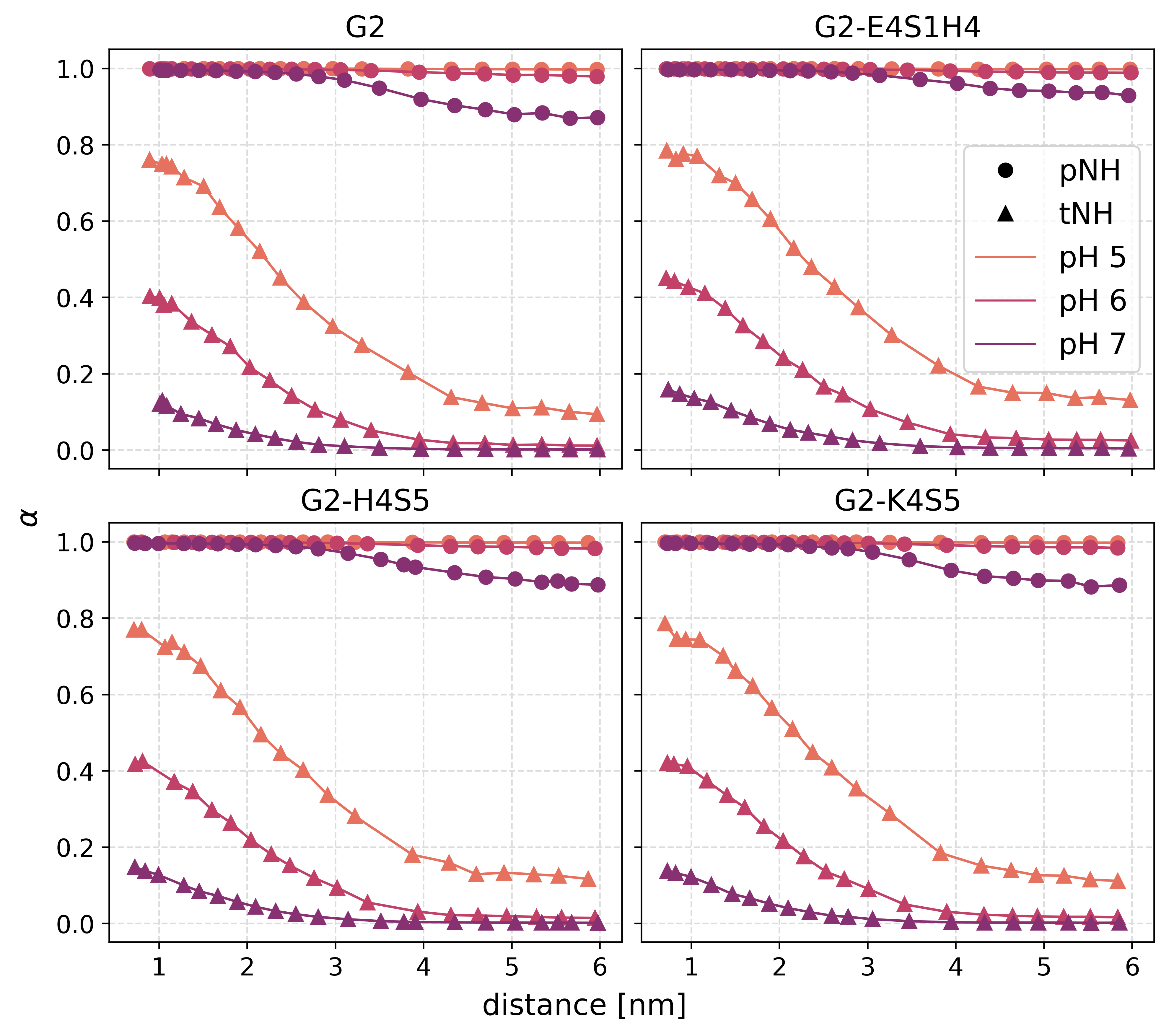}
    \caption{Degree of ionization ($\alpha$) of the primary (pNH, circles) and tertiary (tNH, triangles) amine groups in G2 PAMAM dendrimers and their peptide conjugates as a function of distance from the center of the DNA, across pH range 5-7.}
    \label{fig:alpha_amine}
\end{figure}

\begin{figure}
    \centering
    \includegraphics[width=\linewidth]{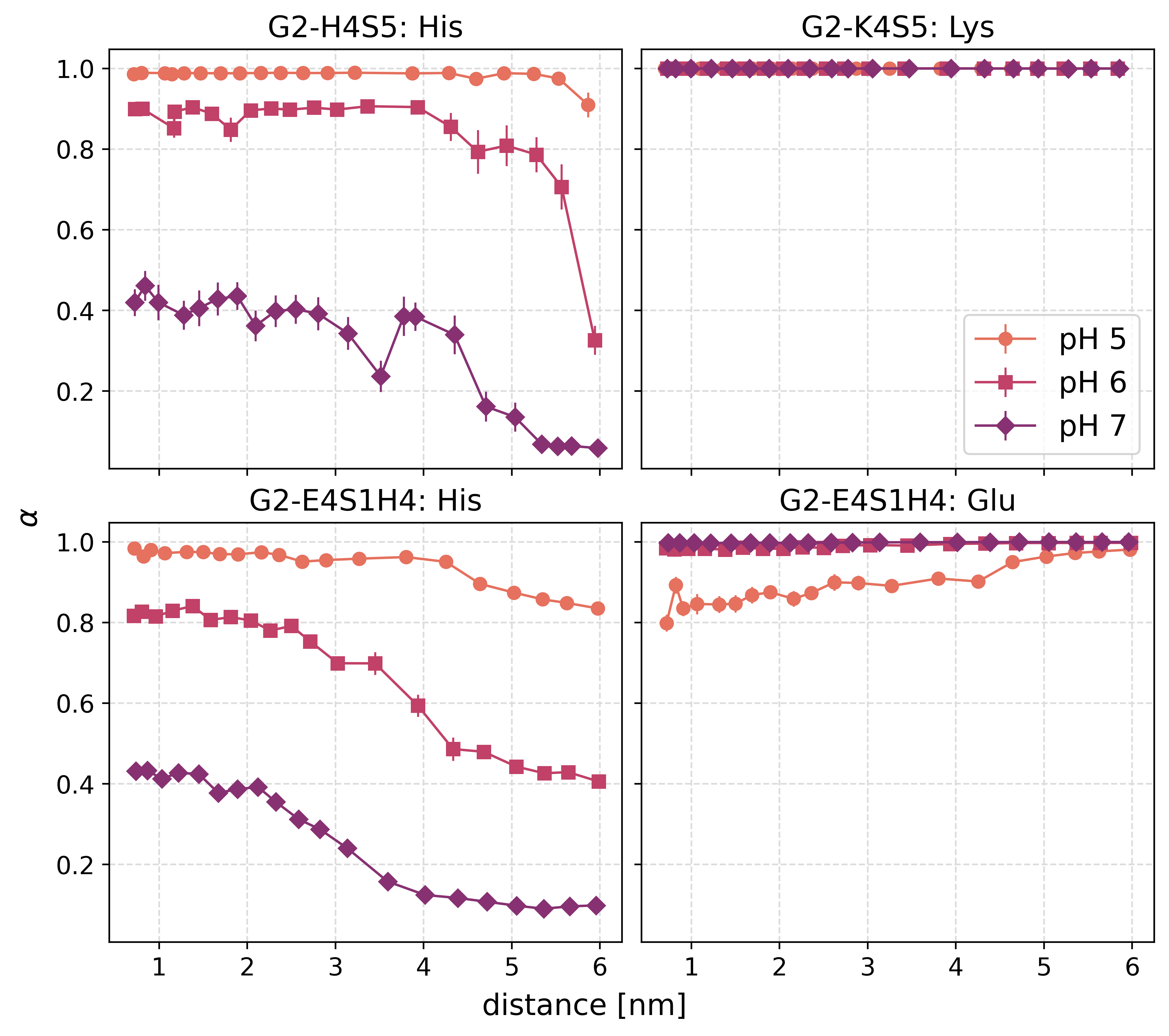}
    \caption{Degree of ionization ($\alpha$) of the ionizable amino acid residues in the peptides of the different G2 PAMAM conjugates as a function of distance from the center of the DNA, across pH range 5-7.}
    \label{fig:alpha_aa}
\end{figure}

The extension and nature of this charge modulation depend on the specific composition of each conjugate and can be rationalized by analyzing the ionization degrees of the various functional groups (\reffig{fig:alpha_amine} and \reffig{fig:alpha_aa}). 
At pH 5 and 6, the increase in net charge of the conjugates upon DNA association is primarily driven by enhanced ionization ($\alpha$) of the tertiary amine groups of G2, while the primary amines remain fully protonated regardless of their distance from DNA. At pH 7, DNA proximity induces only a modest increase in the ionization of tertiary amines, and guarantees full ionization of primary amines. 
Interestingly, in the G2-E4S1H4 conjugate, the primary amines are nearly fully ionized even at pH 7, and the tertiary amines exhibit slightly higher $\alpha$ even at greater distances from the DNA. This behavior is likely due to the glutamic acid residues in the peptides, which are fully deprotonated at pH 7 (\reffig{fig:alpha_aa}), thereby enhancing the ionization of nearby (oppositely charged) amine groups.

Among the conjugates, G2-K4S5 shows the highest net charge across the studied pH range and distances from the DNA (red triangles in \reffig{fig:charge_pH}), as all lysine residues are fully protonated under these conditions.
In contrast, G2-H4S5 shows a similar charge to G2-K4S5 at the lower pH values, but its net charge decreases at pH 7, due to the reduced ionization of the histidine residues (\reffig{fig:alpha_aa}). Consequently, at pH 7, the overall charge of the G2-H4S5 resembles that of the unmodified G2 dendrimer. 

\begin{figure}%[h!tb]
    \centering
    \includegraphics[width=\linewidth]{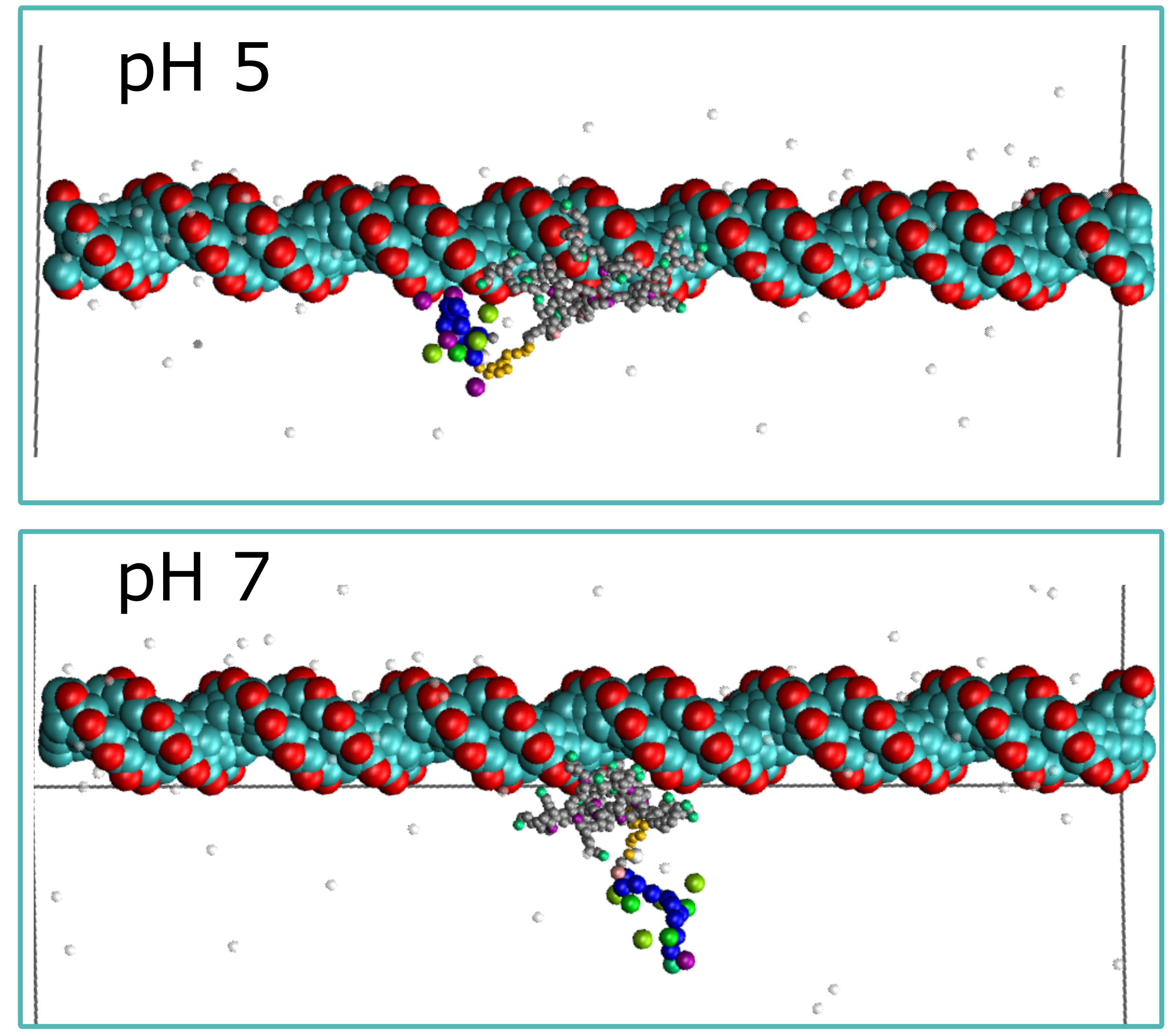}
    \caption{Snapshots of G2-E4S1H4 conjugates and DNA at pH 5 (top) and pH 7 (bottom).}
    \label{fig:snapshot_G2-E4S1H4}
\end{figure}

The net charge of the G2-E4S1H4 conjugate is comparable to that of G2 at pH 5 (\reffig{fig:charge_pH}), due to a slight increase in amine group ionization  (\reffig{fig:alpha_amine}), combined with histidine ionization and reduced ionization of the glutamic acid residues (\reffig{fig:alpha_aa}). As pH increases, the full ionization of glutamic acid residues and the decreasing ionization of histidines, result in a net charge that progressively falls below that of G2. 
Notably, the presence of the glutamic acid residues in G2-E4S1H4 appears to suppress histidine ionization, particularly near the DNA at higher pH. 
This is likely due to electrostatic repulsion between the negatively charged DNA and glutamic acid residues, which causes the peptide to extend away from the DNA, as illustrated in the representative snapshots in \reffig{fig:snapshot_G2-E4S1H4}. While at pH 5, the peptide occasionally interacts with the DNA, at pH 7 it is almost entirely protruding away from the DNA, reducing the influence of DNA-mediated  charge regulation on the histidine residues.

Using computational modeling, we evaluated the potential of mean force (PMF) for the adsorption of G2 dendrimers and their peptide conjugates onto the DNA model.
\begin{figure}
    \centering
    \includegraphics[width=\linewidth]{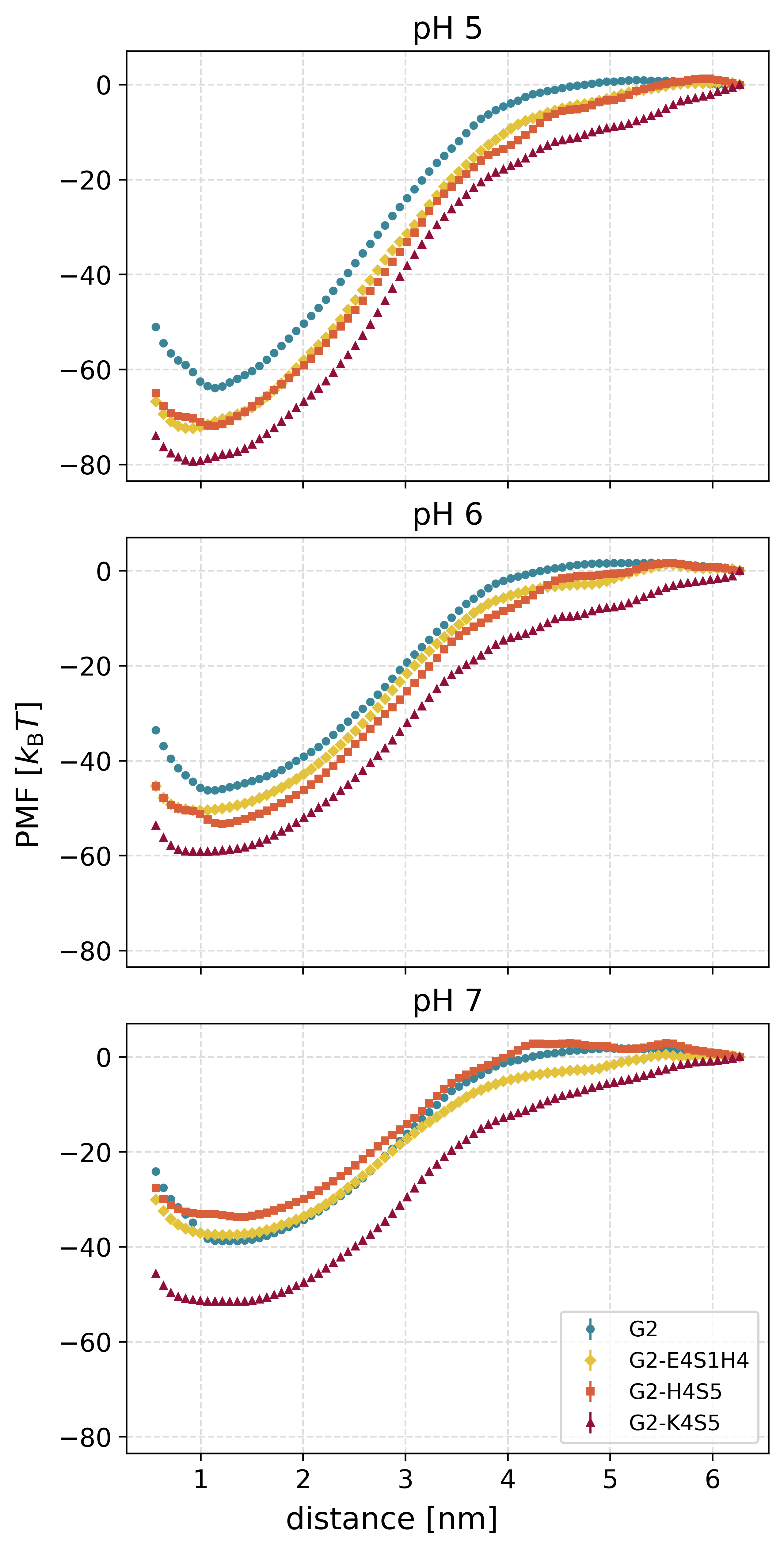}
    \caption{Potential of mean force (PMF) profiles for the adsorption of G2 PAMAM dendrimers and peptide-conjugated dendrimers onto DNA across pH 5–7.}
    \label{fig:PMF_pH}
\end{figure}

As shown in \reffig{fig:PMF_pH}, both the net charge -- modulated by pH and proximity to DNA -- and the spatial distribution of charge within the condensing agents influence their association with DNA.

Focusing first on the unmodified G2 dendrimers (blue symbols), we observe that the depth of the PMF well decreases from $-64k_{\text{B}}T$ at pH 5 to %$-46k_{\text{B}}T$, and to 
$-39k_{\text{B}}T$
at pH 7, reflecting the reduced electrostatic attraction as the dendrimer becomes less charged.
The G2-K4S5 conjugates (red symbols), which carry fully ionized lysine residues across the studied pH range, exhibit the strongest interaction with DNA. These conjugates display both deeper and broader potential wells, the latter attributed to the extended reach of the peptide chains, which enhance interactions even at relatively long distances.

Interestingly, both histidine-containing conjugates, G2-H4S5 and G2-E4S1H4, show similar PMF profiles near the DNA. At pH 5 and 6, their PMF curve falls between those of G2 and G2-K4S5, despite G2-E4S1H4 having a net charge similar to or lower than G2, and G2-H4S5 having a charge comparable to G2-K4S5. At pH 7, the PMFs of G2-H4S5 and G2-E4S1H4 converge with that of G2, even though G2-E4S1H4 has a lower net charge than both G2-H4S5 and G2. This suggests that at this pH, the peptides -- being nearly neutral or negatively charged --  do not significantly contribute to the association with DNA.

These findings are consistent with previous Monte Carlo simulations, which showed that neutral peptides do not significantly affect the binding of PAMAM-peptides conjugates to DNA.~\cite{Dannert2023DNACharge}

\subsection{G2 PAMAM-Peptide Conjugates Exhibit pH-Sensitive DNA Condensation}

\begin{figure}
    \centering
    \includegraphics[width=\linewidth]{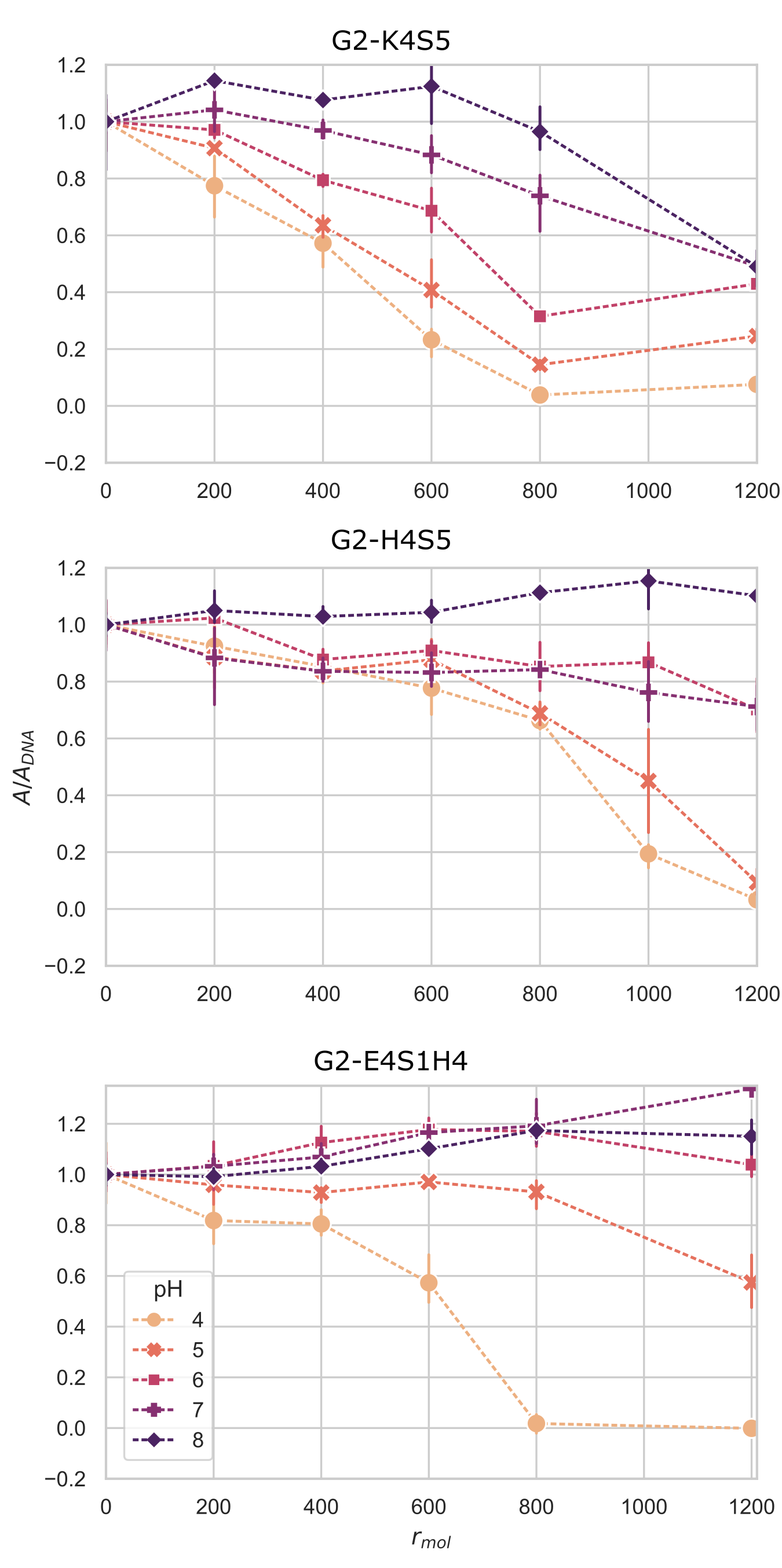}
    \caption{Normalized DNA absorbance at 260 nm following precipitation induced by G2-peptide conjugates at varying molar ratios ($r_{\text{mol}}$) and pH values.}
    \label{fig:precipitation_conjugates}
\end{figure}

Unlike unmodified G2 PAMAM dendrimers, DNA condensation induced by G2-peptide conjugates exhibits a strong dependence on the pH of the buffer solution, as measured using a precipitation assay (\reffig{fig:precipitation_conjugates}). 

Starting with G2-K4S5 (top panel in \reffig{fig:precipitation_conjugates}), it can be seen that the precipitation curve at pH 4 closely resembles that of G2 (\reffig{fig:precipitation_G2}), despite G2  having a lower net charge by approximately $4e$ (\reffig{fig:charge_pH}), and a weaker interaction with DNA (\reffig{fig:PMF_pH}). As the pH increases, the absorption of the supernatant at $\lambda$ = 260 nm also increases, indicating a larger number of DNA molecules in solution after centrifugation. This trend is somewhat unexpected, given that G2-K4S5 maintains a relatively high net charge even at pH 7 -- higher than G2, whose precipitation behavior remains largely  unaffected by pH. 

These findings can be compared to our previous work, where conjugation of peptides containing up to 12 lysine residues to G2, resulted  in a reduced $r_{\rm mol}$ needed for DNA condensation at pH 7.~\cite{Dannert2023DNACharge} In that study, the peptides were shown to stretch along the DNA, occupying the space between the dendrimers. We speculate that while a single G2-K4S5 conjugate may enhance interaction due to its $+4e$ charge, this effect diminishes when multiple conjugates are present. Electrostatic  repulsion between dendrimers adsorbed onto the DNA, combined with the low charge of the peptides, may suppress lysine ionization and promote peptide desorption. 
The conjugation of neutral peptides to G2 has been shown to hinder DNA condensation.~\cite{Dannert2023DNACharge}

Charge regulation is strongest near the \pKa\ of the ionizable groups. Since lysine has a high \pKa (10.4), the observed strong pH-dependence of G2-K4S5 precipitation in the pH 4-8 range may seem counterintuitive. However, this behavior is consistent with the idea that even small changes in local environment or conformation can significantly affect ionization and complex formation 

The observed pH-dependent behavior is also very pronounced in the DNA--G2-H4S5 system, but shifted to higher $r_{\text{mol}}$. Significant DNA condensation is observed at pH 5 and below, provided that the $r_{\text{mol}}$ is sufficiently high. However, condensation efficiency drops sharply at pH values near or above the \pKa\ of histidine.  
An even stronger effect is observed for the G2-E4S1H4 system, fueled by the neutralized peptide at high pH values. It is however unclear why it shows a precipitation profile similar to G2-K4S5 at pH 4.

\begin{figure}
    \centering
    \includegraphics[width=\linewidth]{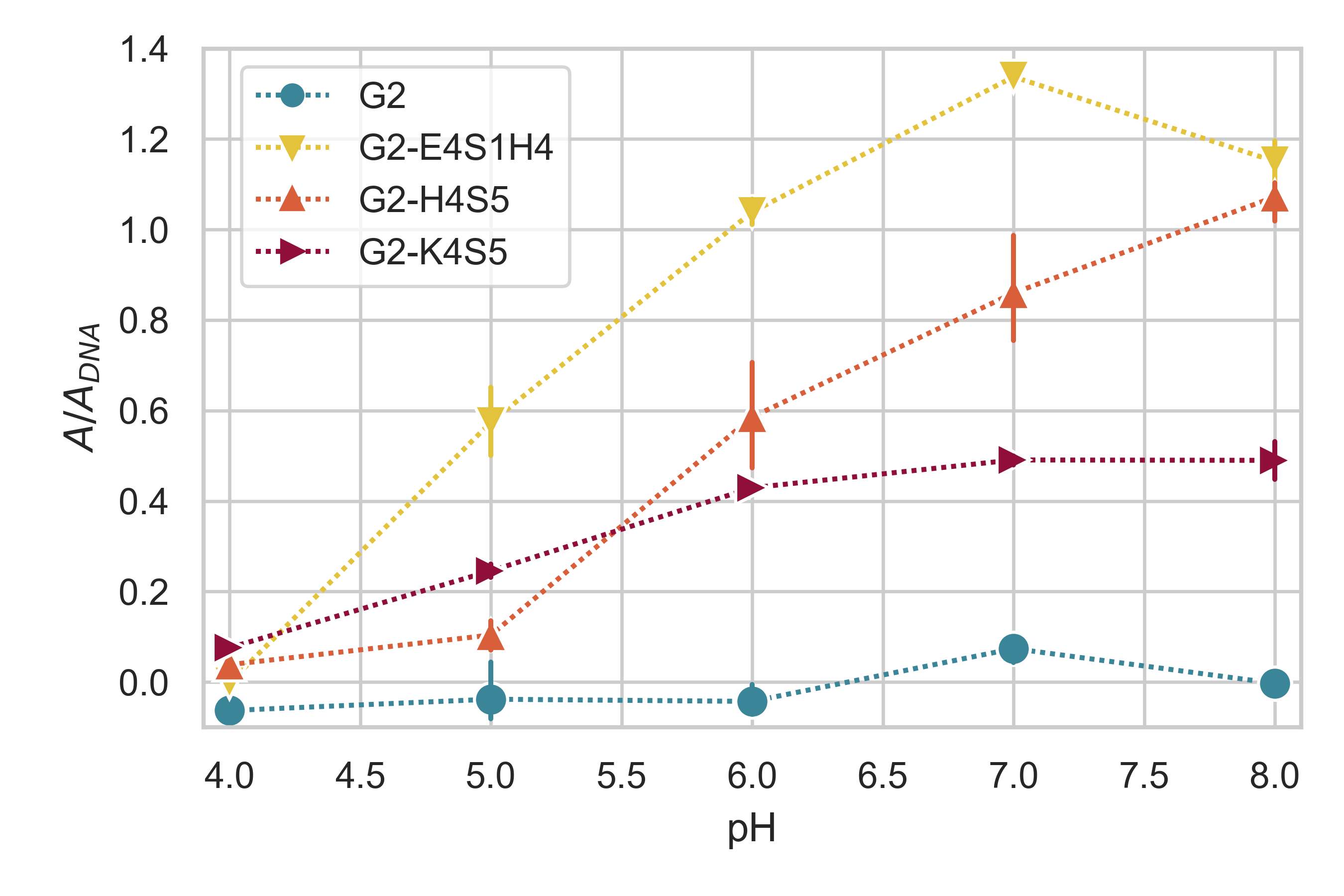}
    \caption{DNA precipitation with G2, G2-H4S5 and G2-K4S5 at $r_{mol}=1200$ with varying pH.}
    \label{fig:precipitation_comparison}
\end{figure}

\reffig{fig:precipitation_comparison} summarizes the precipitation assay results for all systems at $r_{mol}=1200$.
For reference, G2 at pH 7.0 has an average net charge of $+18e$ (when adsorbed to DNA), corresponding to a N/P charge ratio of 1.6. As previously discussed, G2-induced DNA condensation is largely unaffected by pH at this concentration. 
In contrast, G2-K4S5 shows a gradual decrease in DNA condensation with increasing pH, with only $\sim$50\% of DNA  excluded from solution at pH 8. 
The most pronounced pH dependence is observed for the histidine-containing conjugates, G2-H4S5 and G2-E4S1H4. At pH 4, nearly all DNA is condensed, while at pH 8, DNA remains fully in solution. 
This behavior aligns with the steep increase in net charge per conjugate as pH decreases from 8 to 4. In comparison, the charge increase for G2-K4S5 over the same pH range is much smaller, explaining the more moderate pH response.

To summarize, our previous work showed that conjugation of long, highly positively charged peptides to G2 PAMAM dendrimers reduces the amount of dendrimers required to induce DNA condensation.~\cite{Dannert2023DNACharge} In contrast, the present study shows that short peptides with relatively low net charge reduce the condensation efficiency. These findings indicate that peptide conjugation influences DNA condensation through mechanisms beyond a simple increase in the total positive charge of the carrier.

At the level of individual conjugates, the simulations reveal a strong dependence of DNA--conjugate association on both peptide composition and pH. However, these trends do not directly translate into the experimentally observed condensation behavior, highlighting the importance of collective many-body effects during DNA condensation, including interactions between multiple DNA-bound conjugates, charge regulation, and peptide conformational entropy.

Together, these results suggest that short peptide tails primarily regulate the interaction between PAMAM dendrimers and DNA, rather than merely contributing additional charge. 
Depending on their amino acid composition, the peptides modify the pH responsiveness of the conjugates, alter the spatial distribution of charge, and influence how neighboring conjugates can pack along the DNA molecule.

From a gene-delivery perspective, histidine-con\-tai\-ning conjugates are of particular interest because their ionization changes substantially within the pH range encountered during endosomal maturation. The resulting increase in the charge of the DNA-conjugate complexes may enhance their interactions with negatively charged endosomal membranes, a mechanism that has been proposed to contribute to endosomal escape in polycation-based delivery systems.\cite{Bus2018} At the same time, these conjugates were the least efficient DNA-condensing agents at neutral pH, highlighting a trade-off between endosomal responsiveness and nucleic-acid packaging efficiency. Interestingly, lysine-containing conjugates exhibited quantitatively similar pH-dependent changes in DNA condensation at low conjugate concentrations despite the much weaker pH dependence expected from the ionization state of the lysine residues. 

\section{Conclusions}
We combined potentiometric titrations, DNA precipitation assays, and molecular simulations to investigate how the conjugation of short peptides affects the interaction of G2 PAMAM dendrimers with DNA. 

The simulations show that the ionization state of G2 PAMAM dendrimers is strongly influenced by both pH and proximity to DNA. Charge regulation enhances dendrimer protonation near DNA, primarily through the tertiary amine groups, leading to stronger dendrimer–DNA interactions at lower pH. Despite these changes in interaction strength, DNA condensation induced by unmodified G2 PAMAM dendrimers was largely insensitive to pH within the investigated range, consistent with the dendrimers retaining a sufficiently high positive charge under all conditions studied.

The free peptides were found to be insufficiently charged to induce DNA condensation. In contrast, conjugation of a single peptide to G2 substantially altered the behavior of the dendrimers. All peptide-conjugated systems exhibited a pronounced pH dependence of DNA condensation, with condensation efficiency decreasing as the pH increased. This effect was strongest for the histidine-containing conjugates, whose net charge changes markedly within the physiologically relevant pH range.

Molecular simulations further revealed that the interaction strength between the conjugates and DNA depends strongly on both pH and peptide composition. At acidic pH, peptide-containing conjugates generally displayed stronger DNA association than unmodified G2. At neutral pH, however, conjugates containing histidine residues exhibited binding free-energy profiles similar to those of G2, indicating that the peptide tails become nearly neutral and contribute little to the interaction with DNA. Instead, the tails tend to extend away from the DNA surface and remain only weakly involved in binding.

The differences between the single-conjugate PMF calculations and the experimentally observed condensation behavior highlight the importance of collective many-body effects, including conjugate–conjugate repulsion, charge regulation, and peptide conformational entropy, during DNA condensation. Elucidating the role of these collective effects through simulations containing multiple conjugates is the focus of ongoing work.

In summary, conjugation of short peptide tails transforms G2 PAMAM dendrimers from relatively pH-insensitive DNA condensing agents into pH-responsive DNA-binding systems. These modifications increase the pH responsiveness of the conjugates and can lead to a substantial increase in complex charge within the pH range encountered during endosomal maturation, which may enhance interactions between DNA-conjugate complexes and negatively charged endosomal membranes. However, this gain is accompanied by a reduction in DNA condensation efficiency, highlighting a trade-off between endosomal responsiveness and nucleic-acid packaging when designing PAMAM-based gene delivery vectors.

%%%%%%%%%%%%%%%%%%%%%%%%%%%%%%%%%%%%%%%%%%
\section*{Acknowledgments}
We would like to thank Marius Aarsten, Astrid Gaarder Harsheim and Tommy Duy Tran for their contribution to this work during their studies at NTNU. P.K. and R.S.D thank EEA and Norway grants (EHP11BFNU-OVNKM-4-215-01-2022).  The authors acknowledge the computational resources from the High Performance Computing center IDUN at NTNU.

\bibliographystyle{unsrtnat}%{elsarticle-num}
\bibliography{references}

\appendix

\section{Conjugation of PAMAM G2 and peptides}
\label{section:SI_Conjugation_of_PAMAM_G2_and_peptides}
\begin{figure}[h!tb]
    \centering
    \includegraphics[width=\linewidth]{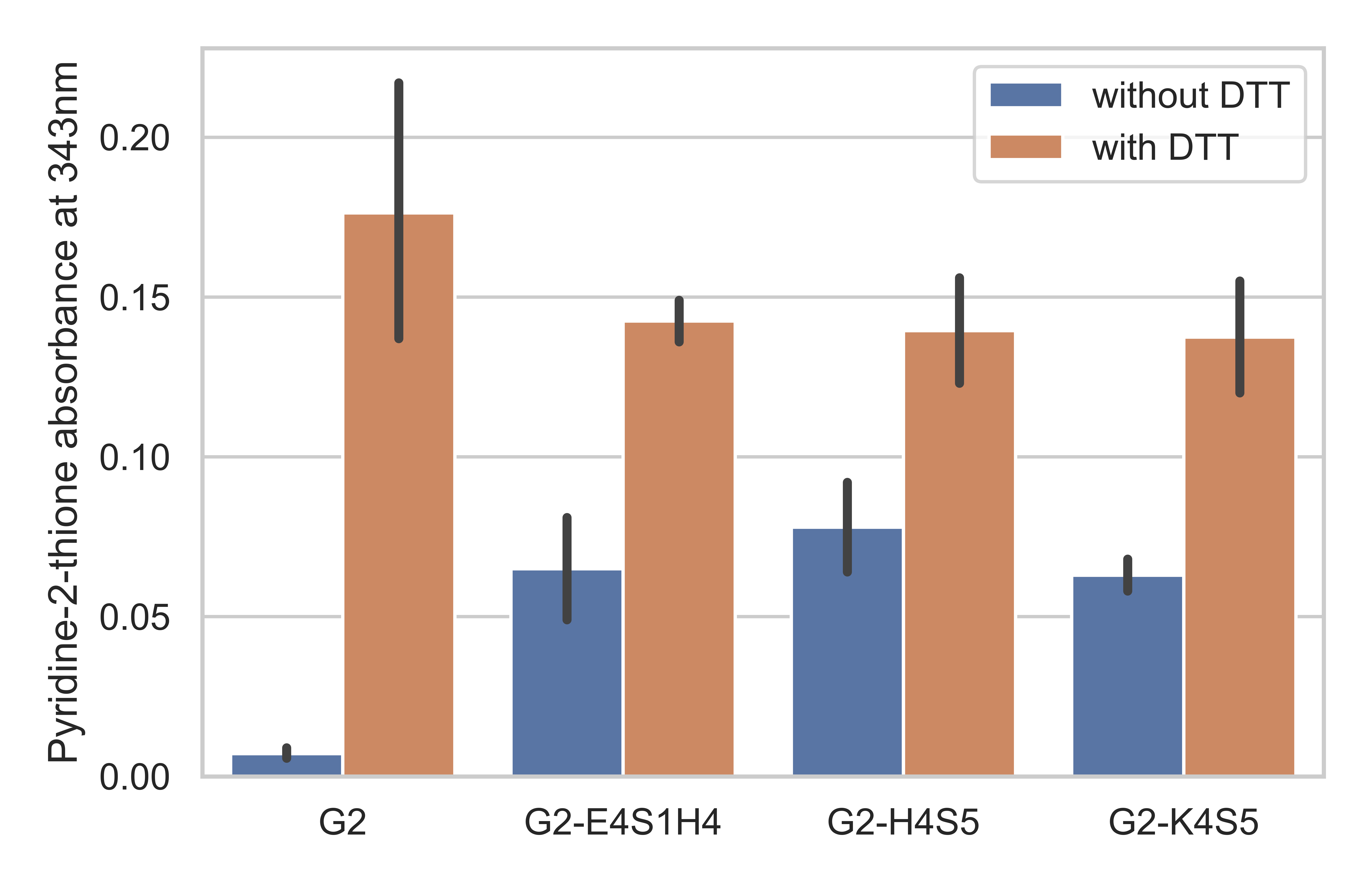}
    \caption{Pyridine-2-thione UV absorbance at 343 nm after conjugating G2 PAMAM dendrimers with Sulfo-LC-SPDP linker before and after addition of DTT (G2) and after conjugating G2-SPDP with peptides before and after addition of DTT.}
    \label{fig:DTT assay}
\end{figure}

\reffig{fig:DTT assay} shows the pyridine-2-thione absorbance at 343 nm measured before and after treatment with DTT. G2 refers to the G2 PAMAM dendrimers modified with a Sulfo-LC-SPDP linker.

\section{Potentiometric titration}
\label{section:SI_Titration}
\subsection{Treatment of titration data}
Potentiometric titration experiments evaluate the solution pH as a function of volume of the titrant (NaOH). To compare the titration data with the simulations and analytical predictions, the charge of the peptides and conjugates must be assessed as a function of pH.

First, the total charge of all molecules present in the solution $Z_{\rm tot}$ needs to be estimated, with
\begin{equation}
    Z_{\rm tot} = z^{\text{OH}^-}+z^{\text{H}^+}+z^{\text{Cl}^-}+z^{\text{Na}^+}      ,
\end{equation}
where $z^{\text{OH}^-}$, $z^{\text{H}^+}$, $z^{\text{Cl}^-}$ and $z^{\text{Na}^+}$ are the total amount of charge due to protons, hydroxide, chloride anions and sodium cations, respectively. 

The amount of charge due to protons and hydroxide can be calculated from the measured pH using 
\begin{equation}
    z^{\text{H}^+}=e10^{-\text{pH}}cV_{\rm tot}
\end{equation}
and
\begin{equation}
    z^{\text{OH}^-}=-e10^{\text{pH}-pK_w}cV_{\rm tot} \,,
\end{equation}
where $c$ is the reference concentration of \SI{1}{\mol/\kilo\gram}, $V_{\rm tot}$ is the total volume of the solution (including the added titrant), and ${\rm p}K_w=\log_{10}K_w$ is the self-ionization equilibrium constant of water. 
The number of charges due to the titrant NaOH and HCl stock are given by 
\begin{equation}
z^{\text{Na}^+}=ec_{\text{NaOH}}V_{\text{NaOH}}
\end{equation}
and
\begin{equation} 
    z^{\text{Cl}^-}=-ec_{\text{HCl}}V_{0}   \,,
\end{equation}
where $c_{\text{NaOH}}$ and $c_{\text{HCl}}$ are the concentration of the NaOH and HCl stock solutions respectively. $V_{\text{NaOH}}$ is the added volume of the titrant NaOH solution and $V_0$ is the initial volume of HCl stock solution.

Since our samples are polydisperse and contain multiple polymeric units with different monomers (dendrimers with peptides attached), it is challenging to directly assess the charge of each molecule, or find reasonable monomeric units to calculate the charge per monomer, as has been shown to be efficient previously.~\cite{Pineda2024ChargePolyelectrolytes}

Due to the uncertainty of the size of each sample molecule and possible unaccounted charge $Z_{\text{extra}}$, the titration curves were anchored to the simulation data.
The anchoring points were chosen to be the charge at pH 3 ($z_{\text{sim},3})$ and pH 10 ($z_{\text{sim},10})$. The scaling factor was then calculated using
\begin{equation}
    x =  \bigg|\frac{z_{\text{sim},3}-z_{\text{sim},10}}{z_{\text{exp},3}-z_{\text{exp},10}}\bigg|
\end{equation}
where $z_{\text{exp},3}$ and $z_{\text{exp},10}$ are the experimental charge at pH 3 and pH 10, respectively. 
The shift was then calculated by
\begin{equation}
    y = z_{\text{sim},10}x+z_{\text{exp},10}\,.
\end{equation}
The experimental charge values were then scaled and shifted using
\begin{equation}
    z = z_0x+y \,,
\end{equation}
where $z$ is the processed charge and $z_0$ is the unshifted, measured charge at a given pH. 

The potentiometric titration data was compared to the analytical solution calculated with the Hendersson-Hasselbach equation
\begin{equation}
    \text{pH} = \text{p}K_{\text{a}} + \text{log}\biggl(\frac{[\text{A}]}{[\text{HA}]}\biggr)   \,.
    \label{eq:HH}
\end{equation}

\section{Additional details on molecular simulations}
{\label{sec:MolSims}}
\subsection{Potential energy of the system} {\label{sec:pot_energy}}
The coarse-grained model includes bounded, steric, and electrostatic interactions. 
The total potential energy is expressed as
\begin{equation}
	U = U_{\mathrm{bond}} + U_{\mathrm{LJ}} + U_{\mathrm{elec}}.
\end{equation}

The monomers were bonded using a harmonic potential given by
\begin{equation}
U_\text{bond}(r)=\frac{1}{2}k(r-r_o)^2    
\end{equation}
with the stiffness constant $k=560 \,k_\text{B}T/\sigma^{2}$ and $\sigma=1.5$ Å. $r_0$ refers to the equilibrium length, whose values for the different bonds in the models are shown in Table \ref{tab:model_parm}, together with the bead diameters.

The DNA was represented as a stationary double helix with 6 turns. In addition, counter-ions were added to the system to ensure electroneutrality. 

\begin{table}[htb]
\centering
\caption{Parameters (bead diameter and $r_0$) used in the coarse-grained models of G2 PAMAM, linker and peptides.}

\begin{tabular}{c|c|c}
    \toprule
    Group & Bead [Å] & $r_0$ [Å] \\
    \midrule
    PAMAM  & 1.5 & 1.54\\
    linker & 1.5 & 1.54\\
    disulfide bond (linker - peptide) &- & 1.98\\
    peptide backbone & 3.02 & 3.1\\
    peptide N-terminus & 1.50 & 3.00 \\
    peptide C-terminus& 3.00 & 3.10\\
    Glycine&  3.02 & 3.64\\
    Cysteine & 2.41 & 4.62\\
    Serine & 2.22 & 2.69 \\
    Histidine & 3.06 & 5.53\\
    Lysine & 3.40 & 6.02\\
    Glutamic acid & 3.18 & 6.10 \\
    \bottomrule
\end{tabular}
\label{tab:model_parm}
\end{table}

Steric non-bonded interactions between all beads (except between the fixed DNA beads)  are modeled using a pairwise augmented Lennard--Jones (LJ) potential,\cite{beyer2025charge}
\begin{equation}
	\begin{split}
		U_{\mathrm{LJ}}(r) =
		\begin{cases}
			\infty,
			 & r < r_{\mathrm{off}},                  \\[4pt]
			4\epsilon
			\left[
				\left(
				\displaystyle\frac{\sigma}{r-r_{\mathrm{off}}}
				\right)^{12}
				-
				\left(
				\displaystyle\frac{\sigma}{r-r_{\mathrm{off}}}
				\right)^6
				\right]
			-C,
			 & r_{\mathrm{off}} < r < r_{\mathrm{c}}, \\[10pt]
			0,
			 & r > r_{\mathrm{c}},
		\end{cases}
	\end{split}
	\label{eq:lj}
\end{equation}
where $r$ is the center-to-center distance between particles $i$ and $j$, $\sigma$ defines the interaction range, $\epsilon$ sets the energy scale, $r_{\mathrm{c}}$ is the cutoff distance, and $r_{\mathrm{off}}$ is a size-dependent offset.
The constant
\begin{equation}
	C =
	4\epsilon
	\left[
		\left(
		\frac{\sigma}{r_{\mathrm{cut}}}
		\right)^{12}
		-
		\left(
		\frac{\sigma}{r_{\mathrm{cut}}}
		\right)^6
		\right]
\end{equation}
ensures continuity of the potential at the cutoff, with $
	r_{\mathrm{cut}} = r_{\mathrm{c}} - r_{\mathrm{off}}$.

We set $\sigma=1$ for all particle pairs, such that the shape of the LJ interaction is independent of particle size.
Differences in particle diameter are instead incorporated through the offset
\begin{equation}
	r_{\mathrm{off}}
	=
	\frac{1}{2}(d_i+d_j)-\sigma,
	\label{eq:offset}
\end{equation}
where $d_i$ and $d_j$ are the diameters of the interacting particles.
This construction ensures that the steric contact distance reflects the physical particle sizes.
We further set $r_{\mathrm{c}} = 2^{1/6}\sigma + r_{\mathrm{off}}$ and $\epsilon = k_\mathrm{B}T = 1$, which defines the unit of energy in the reduced-unit system.
With this choice of parameters, Eq.~\ref{eq:lj} reduces to a purely repulsive Weeks--Chandler--Andersen (WCA) potential irrespective of particle size.

Electrostatic interactions were described using a screened Coulomb potential using a Debye-H\"uckel approximation,
\begin{equation}
 U(r_{i,j})= \frac{q_iq_j}{4\pi \epsilon_0\epsilon_r}\frac{\text{exp}(-\kappa r)}{r} \text{  for  }r<r_\text{cut},
 \label{eq:S_Coulomb}
\end{equation}
where $q_i$ is the charge of particle $i$, $\epsilon_0$ and $\epsilon_r$ are the vacuum permittivity and relative permittivity respectively. The solvent is treated implicitly through the dielectric constant of water, $\epsilon_\mathrm{r}=78.5$, at $T=\qty{298.15}{K}$.
 The cutoff was set to be half of the height of the box. The Debye length is given by
\begin{equation}
    \kappa^{-1}=\frac{0.304}{\sqrt{I}},
\end{equation} 
where $I$ is the ionic strength of the solution. 

The ionic strength of the system was set to \SI{17}{\milli\molar}. 
The simulation box was setup with periodic boundary conditions in a square prism.
The system’s energy was relaxed with 5000 steepest descent integrator steps to eliminate any overlap between particles.

\subsection{Langevin Dynamics}

Configurational sampling was performed by integrating the Langevin equation of motion,
\begin{equation}
	m \frac{{\rm d} \mathbf{v}}{{\rm d} t}=
	- \gamma \mathbf{v}
	+ \sqrt{2\gamma \,k_\mathrm{B}T}\,\boldsymbol{\xi}
	- \nabla \mathbf{U}(\mathbf{x}),
	\label{eq:langevin}
\end{equation}
where $m$ and $\mathbf{v}$ are the particle mass and velocity, respectively, $\gamma$ is the damping (Stokes) coefficient, and $\mathbf{U}(\mathbf{x})$ is the total potential energy arising from the molecular interactions described in Sec.~\ref{sec:pot_energy}.
The thermal energy is given by $k_\mathrm{B}T$, where $k_\mathrm{B}T$ is the Boltzmann constant and $T$ is the temperature.

The stochastic term $\boldsymbol{\xi}$ represents Gaussian white noise originating from collisions with the implicit solvent and satisfies
\begin{align}
	\left< \boldsymbol{\xi}(t) \right>                          & = 0, \label{eq:xi_mean} \\
	\left< \boldsymbol{\xi}(t)\cdot\boldsymbol{\xi}(t') \right> & = d\,\delta(t-t'),
	\label{eq:xi_var}
\end{align}
where $d=3$ is the dimensionality of the system, $\delta$ is the Dirac delta function, and $\left< \cdots \right>$ denotes an ensemble average.
Together, the dissipative and stochastic terms reproduce the Brownian motion of particles in an implicit solvent and ensure canonical sampling at equilibrium.

Simulations were performed in reduced units with thermal energy $k_\mathrm{B}T=\epsilon=1$, where $\epsilon$ defines the unit of energy.
The equations of motion were integrated using a Velocity--Verlet scheme with time step $\Delta t = 0.005\tau$ and damping coefficient $\gamma = 10\,\tau^{-1}$, where $\tau = \sigma\sqrt{m/\epsilon}$ is the unit of time.
These parameters were chosen to ensure stable and efficient sampling of configurational space.
The particle mass was set to $m=1$, as it does not affect equilibrium thermodynamic properties.

\subsection{Constant pH Monte Carlo}

Titration Monte Carlo moves in the constant-pH ensemble were used to simulate the dissociation reactions of weak acid and base groups, represented as
\begin{equation}
  \text{HA}  \; {\rightleftharpoons{}} \;\text{A}^{-}+\text{H}^{+} ,
\end{equation}
and 
\begin{equation}
  \text{BH}^+ \; {\rightleftharpoons{}} \;\text{B}+\text{H}^{+} .
\end{equation}
The MC steps involved the protonation or deprotonation of ionizable beads, while simultaneously inserting or deleting a counterion in order to assure electroneutrality. The probability of accepting such a MC trial move is given by~\cite{Reed1992MontePolyelectrolytes}
\begin{equation}
    P_{\text{cpH}} = \text{min}\Bigl[1,\text{exp}\Bigl(-\Delta U/k_\text{B}T+\zeta \text{ln}(10)(\text{pH}-\text{p}K_\text{a})\Bigr)\Bigr],
\end{equation}
where $\Delta U$ is the change in potential energy, $\zeta=\pm1$ for the forward or reverse direction of the reaction, and ${\rm p}K_{\rm a} = -\log K_{\rm a}$.

Following the Addition of a Cation Procedure (ACP) \cite{labbez07b}, protonation and deprotonation moves are coupled to the creation or deletion of a cation in order to preserve electroneutrality.
The activity coefficient of the cation, $\gamma_{\mathrm{Na}^+}$, is accounted for a posteriori by shifting the pH scale, according to \cite{beyer25a}
\begin{equation}
	{\rm pH} = {\rm pH}^{\mathrm{input}} + {\rm p}\gamma_{\mathrm{Na}^+},
\end{equation}
where ${\rm p}\gamma_{\mathrm{Na}^+} = -\log \gamma_{\mathrm{Na}^+}$.
The activity coefficient is estimated using the semi-empirical Davies equation:
\begin{equation}
	p\gamma_{\mathrm{Na}^+} = A \left(
	\frac{\sqrt{I/c^{\ominus}}}{1 + \sqrt{I/c^{\ominus}}}
	- C \frac{I}{c^{\ominus}}
	\right),
\end{equation}
where $A \approx 0.509$ and $C = 0.3$ are empirical parameters for water at 298 K, $I$ is the ionic strength, and $c^{\ominus} = 1~\mathrm{mol\,kg^{-1}}$ is the standard concentration.

\subsection{Radial distribution function}
\begin{figure}
    \centering
    \includegraphics[width=\linewidth]{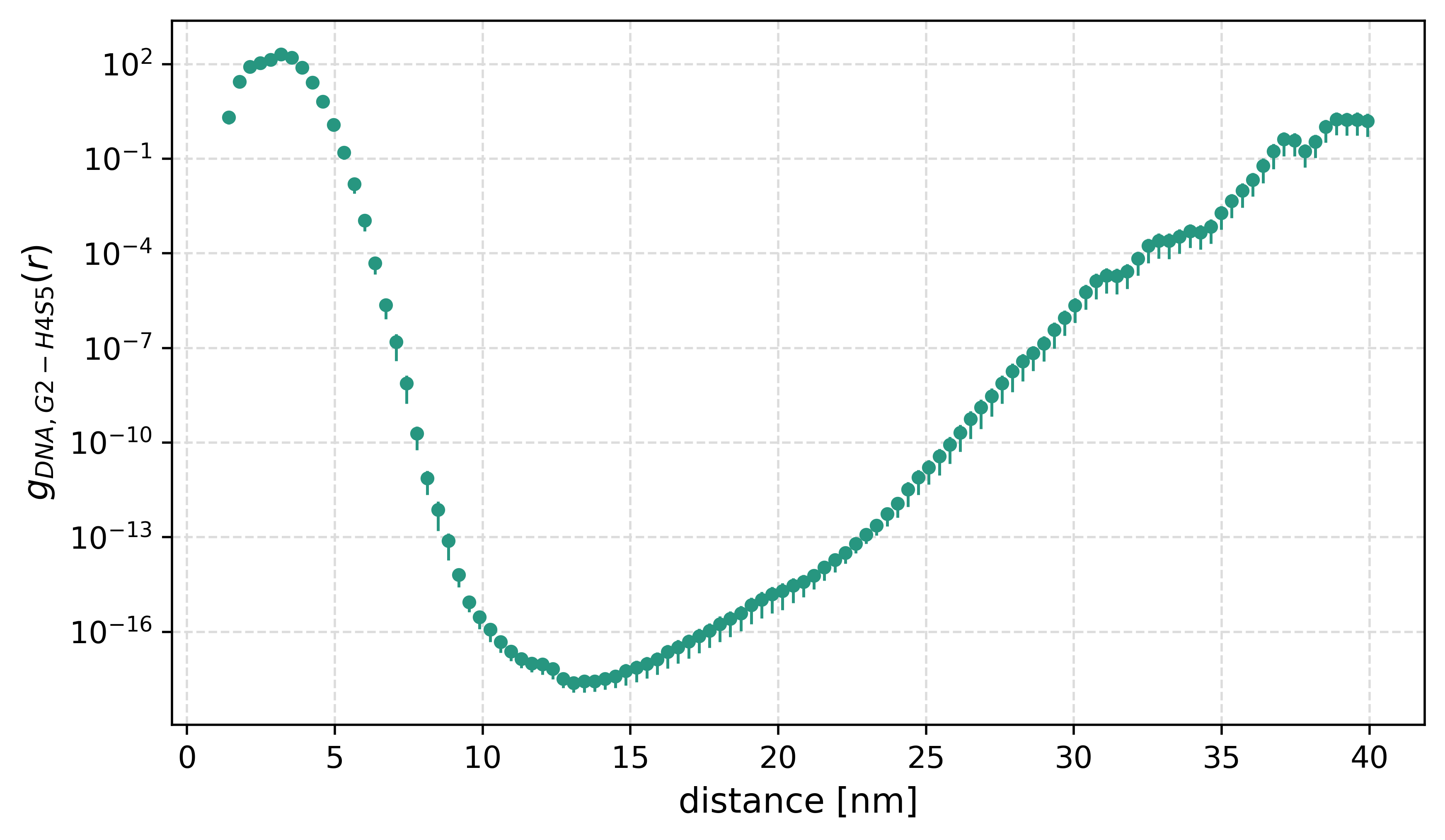}
    \caption{Radial distribution function $g(r)$ of G2-H4S5 as a function of distance from the center of the DNA at pH 7.}
    \label{fig:SI_RDF_H4S5}
\end{figure}
\reffig{fig:SI_RDF_H4S5} shows the radial distribution function of the central beads of the dendrimers in G2-H4S5 conjugates at pH 7 as a function of the distance from the DNA. After a larger peak, there is a depletion zone starting at approximately \SI{5}{\nano\meter} from the center of the DNA. The bulk concentration is reached at approximately \SI{37}{\nano\meter}. This simulation was used to verify that the simulation box was sufficiently large. In order to reduce computational costs, all other umbrella sampling was restricted to \SI{6}{\nano\meter}.

\subsection{Umbrella sampling}
Umbrella sampling was used to calculate free energy profiles of the dendrimer and conjugates as a function of distance from the DNA.The core of the dendrimer was anchored to a phantom particle in the middle of the DNA using a harmonic potential 
\begin{equation}
    U_\text{bias}(\eta) = \frac{1}{2}k(\eta-\eta_\text{0})^2 \,,
\end{equation}
where $\eta$ is the reaction coordinate, $\eta_0$ is the reference point, and $k$ is the strength of the potential that was set to $k=1\,k_\text{B}T/\sigma^2$. The movement of the anchored bead was constrained along the $z$-direction to ensure that the calculated distance from the DNA remained perpendicular to its helical axis.

The weighted histogram analysis method (WHAM)~\cite{Souaille2001ExtensionCalculations} was used to calculate the radial distribution function and potential of mean force.

\section{Precipitation Assay Peptides}
\label{section:Prec_peptides}
\begin{figure}[h!tb]
    \centering
    \includegraphics[width=\linewidth]{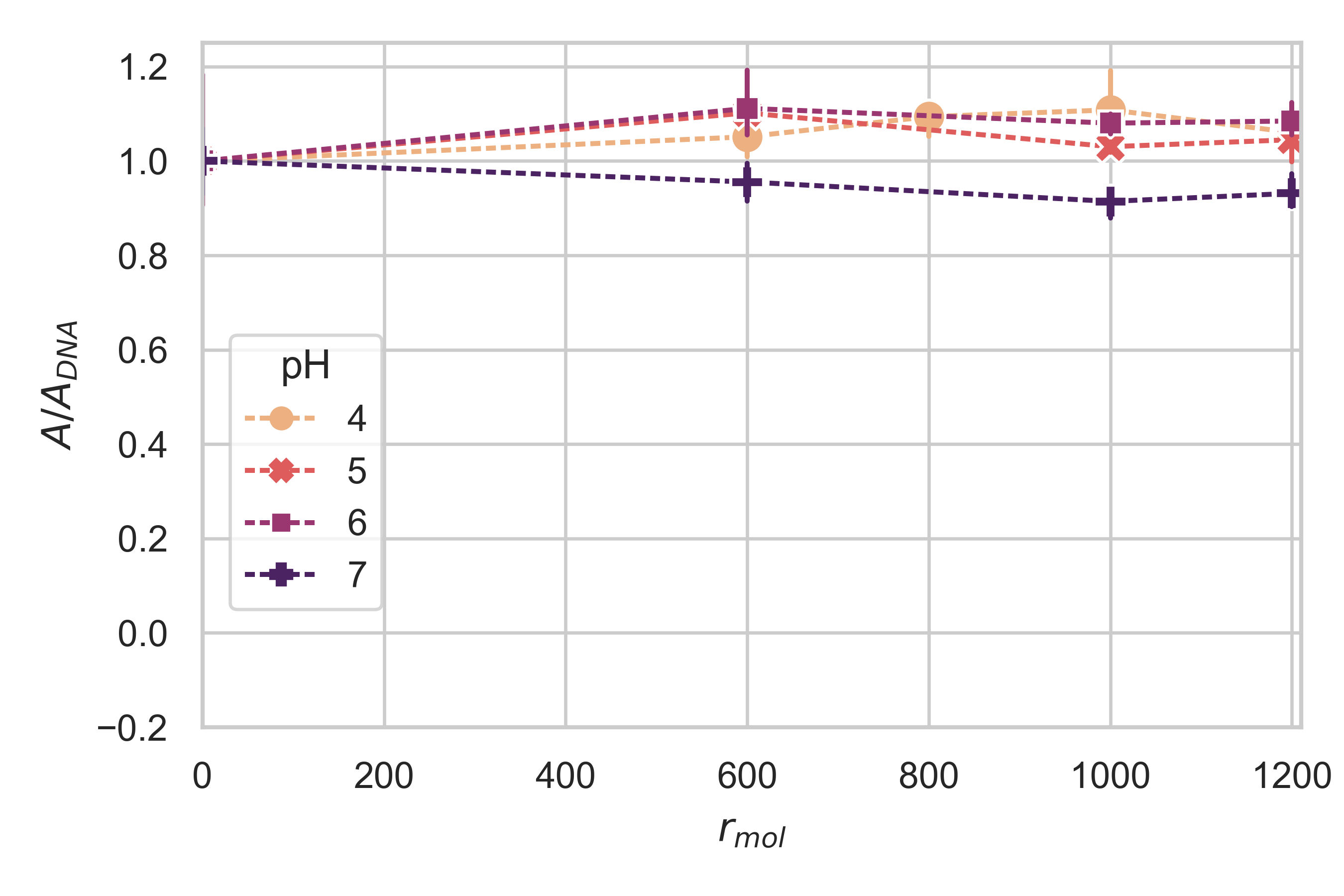}
    \caption{Normalized DNA absorbance as a function of $r_{\text{mol}}$ of K4S5 peptides at varying pH.}
    \label{fig:precipitation_K4S5}
\end{figure}

\reffig{fig:precipitation_K4S5} shows the results of the precipitation assay using K4S5 peptides at varying $r_{\text{mol}}$ and pH. No decrease in UV absorbance of free DNA is visible, independent of pH and $r_{\text{mol}}$ in the investigated range.

\section{Phosphorus-31 NMR spectra}
\label{section:PNMR}

 \subsection{NMR spectra}
 
\begin{figure*}[h!]
    \centering
    \includegraphics[width=0.48\linewidth]{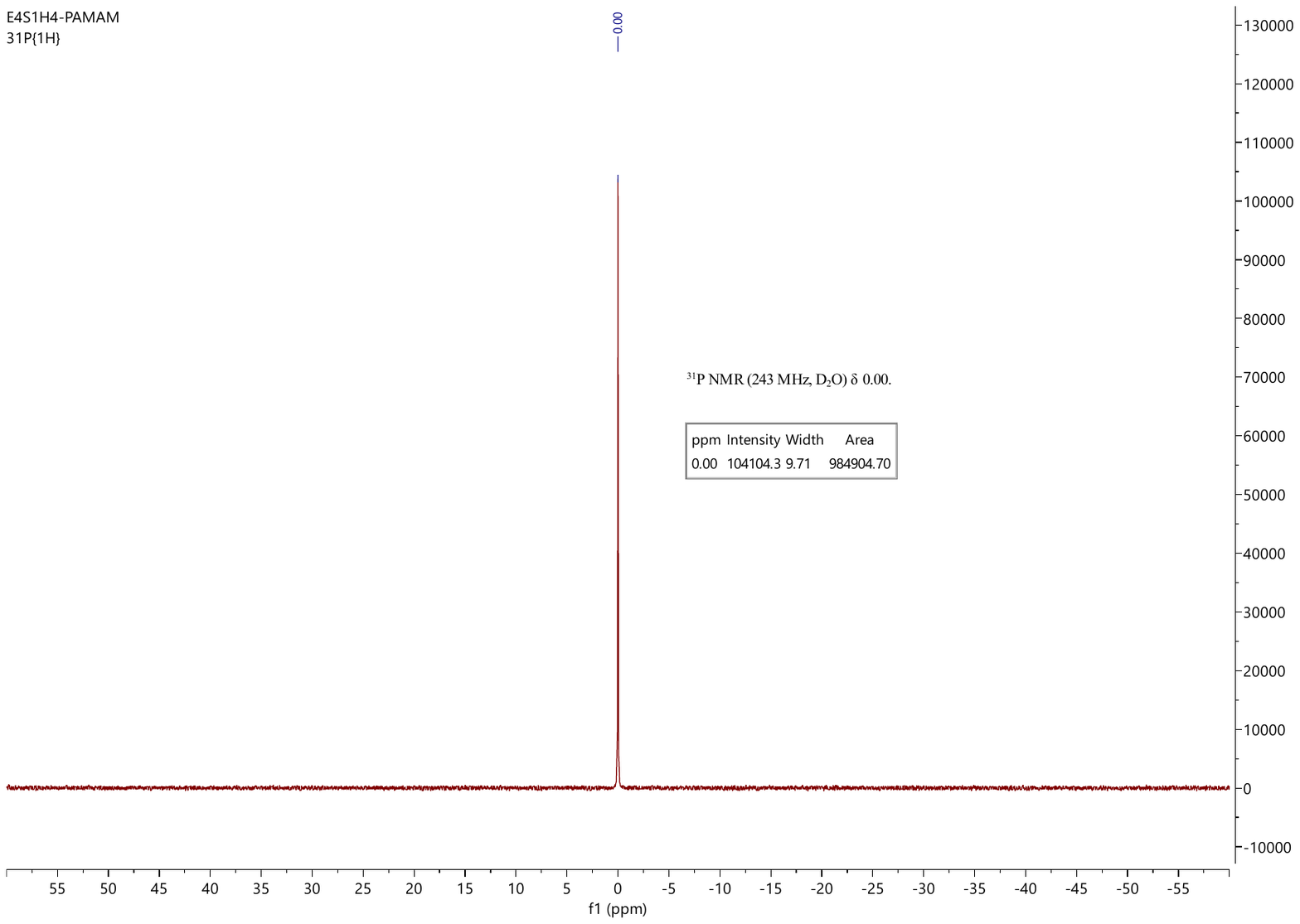} 
    \includegraphics[width=0.48\linewidth]{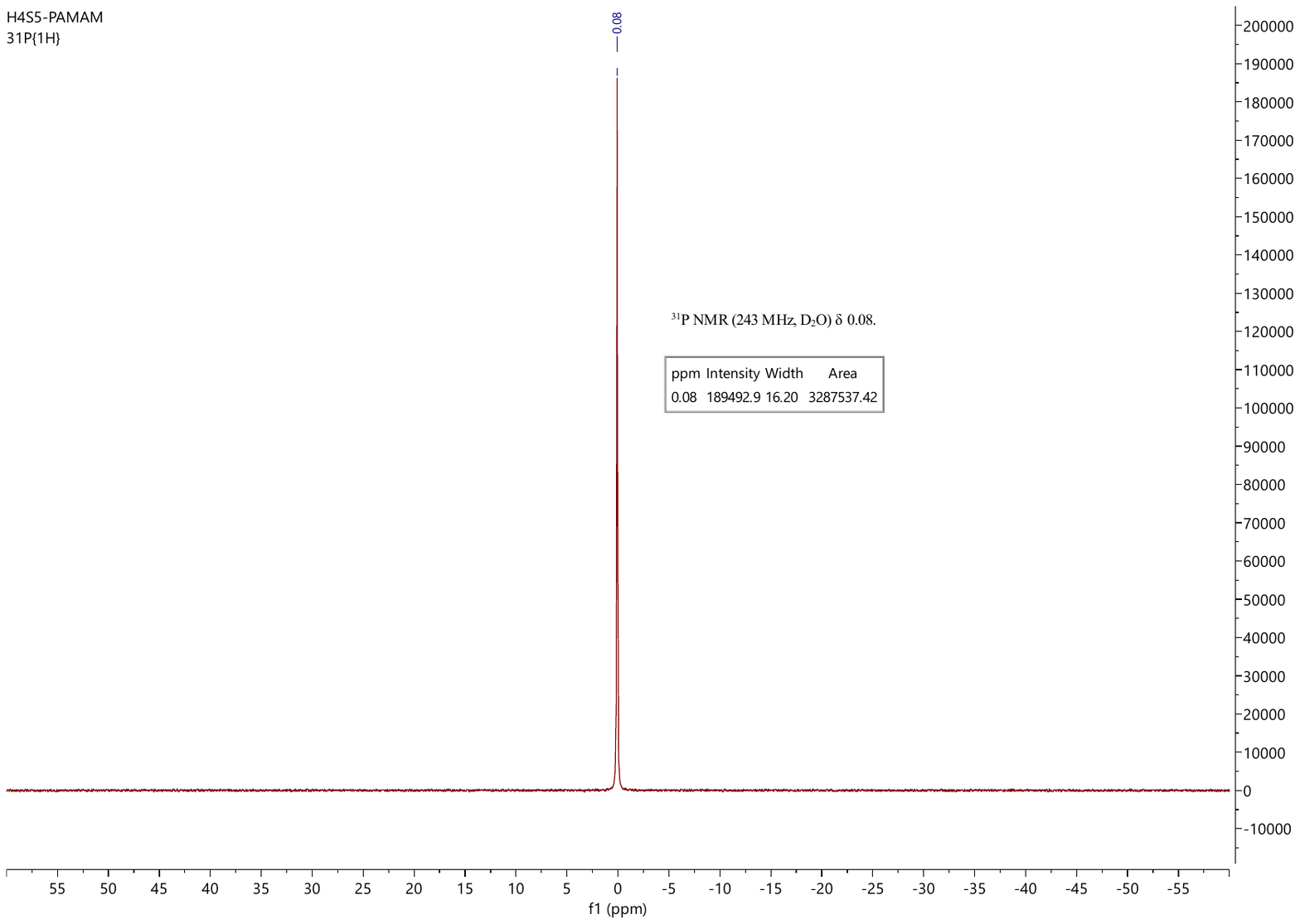} \\
    \includegraphics[width=0.48\linewidth]{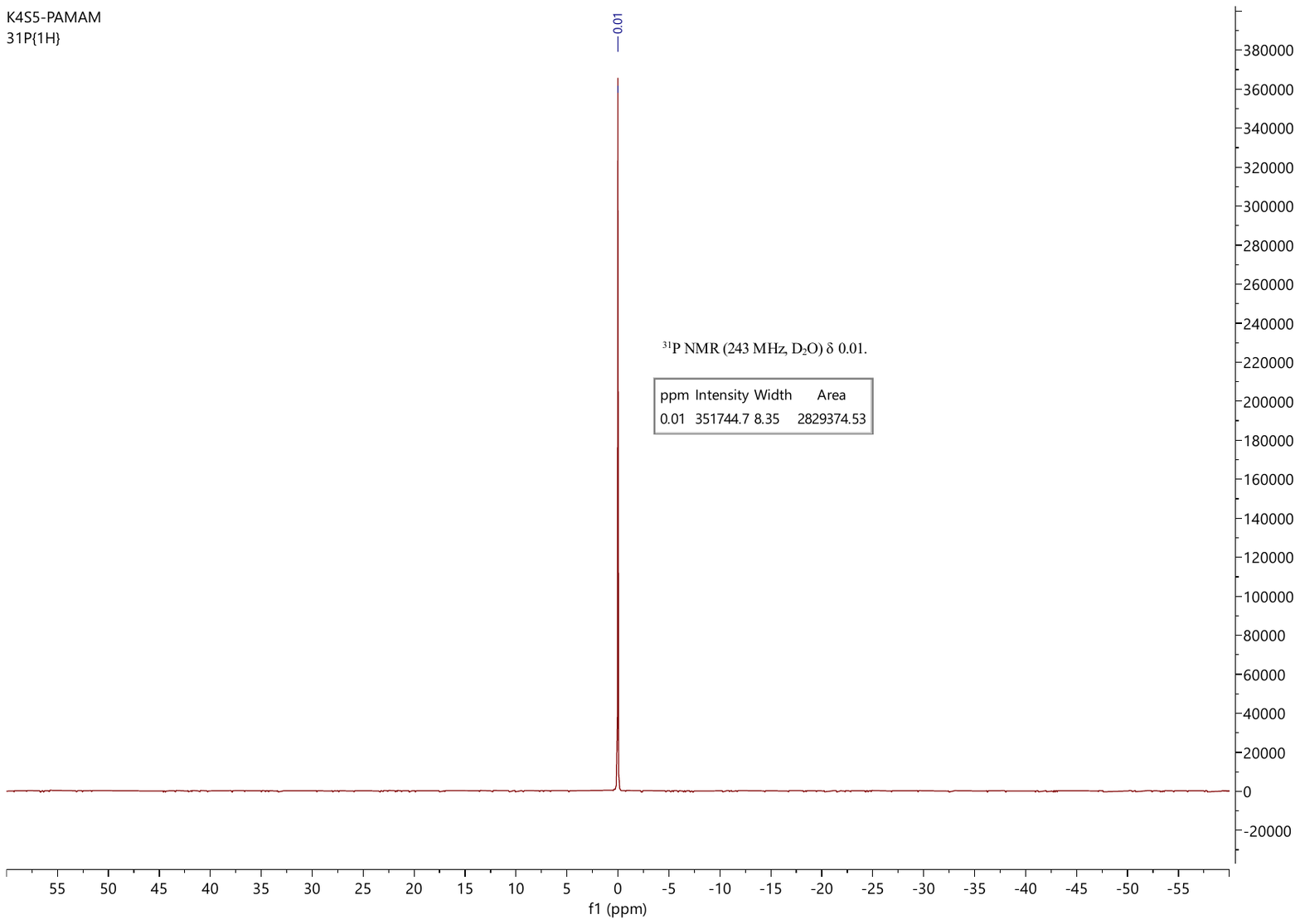} 
    \includegraphics[width=0.48\linewidth]{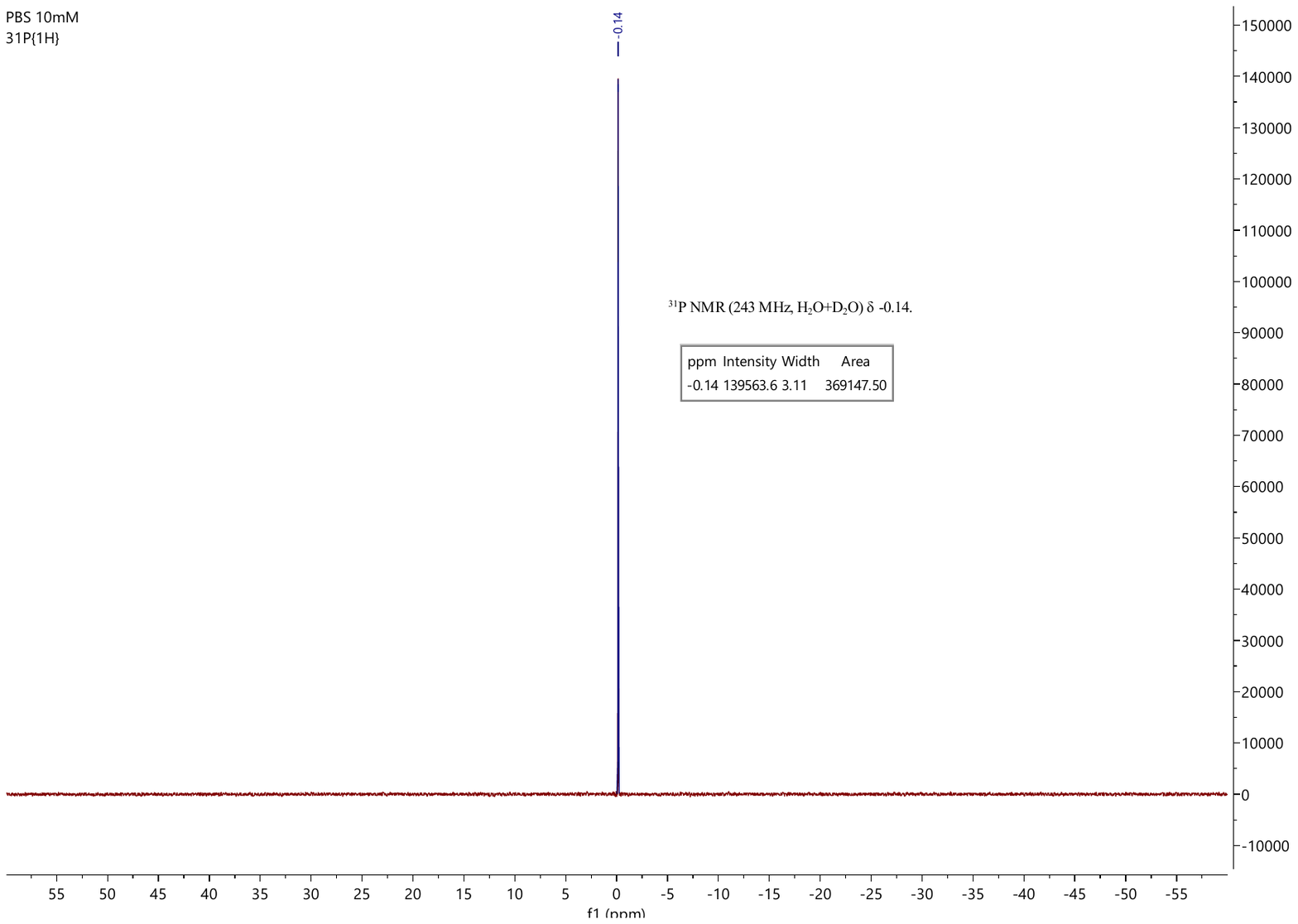} 
     \caption{$^{31}$P NMR spectra of (from top to bottom and left to right): G2-E4S1H4, G2-H4S5, G2-K4S5, and a \SI{10}{\milli\molar} PBS solution. The peak in the G2-conjugate solutions confirms the presence of phosphate in the samples after dialysis.} 
    \label{fig:PNMR}
\end{figure*}

$^{31}$P NMR spectra confirming the presence of phosphorus in the three G2-peptide conjugate samples, and the \SI{10}{\milli\molar} PBS reference are shown in \reffig{fig:PNMR}.

\subsection{Quantification of phosphorus by $^{31}$P NMR}

The phosphorus concentration in each sample was determined by comparingthe $^{31}$P NMR peak area to that of a \SI{10}{\milli\molar} PBS reference solution. 
Assuming a linear relationship between peak area and concentration, the phosphorus concentration in each sample was calculated as:
\begin{equation}
    C_{\mathrm{P}} = C_{\mathrm{PBS}} \cdot 
    \frac{A_{\mathrm{sample}}}{A_{\mathrm{PBS}}},
    \label{eq:NMR_quantification}
\end{equation}
\noindent where $C_{\mathrm{PBS}} = \SI{10}{\milli\molar}$, and $A_{\mathrm{sample}}$ and $A_{\mathrm{PBS}}$ are the integrated peak areas of the sample and reference, respectively. 
The number of moles of phosphorus was obtained from the calculated concentration and the sample volume of approximately \SI{0.6}{\milli\liter}, according to
\begin{equation}
    n_{\mathrm{P}} = C_{\mathrm{P}} \times V_{\mathrm{sample}}
    \label{eq:NMR_moles}
\end{equation}

The integrated peak areas, calculated phosphorus concentrations, and molar amounts are summarized in Table \ref{tab:NMR_areas}. 
The molar ratio between phosphorus and each G2-peptide conjugate is reported in Table \ref{tab:NMR_ratio}.

\begin{table}[h!]
\centering
\caption{Integrated $^{31}$P NMR peak areas, calculated phosphorus concentrations, and molar amounts for each sample. 
The reference PBS concentration was \SI{10}{\milli\molar} and all conjugate sample volumes were \SI{0.6}{\milli\liter}.}
\label{tab:NMR_areas}
\begin{tabular}{lccc}
\hline
\textbf{Sample} & \textbf{Peak Area} & 
\textbf{$C_{\mathrm{P}}$ (mM)} & 
\textbf{$n_{\mathrm{P}}$ (\si{\micro\mole})} \\
\hline
PBS \SI{10}{\milli\molar} & 369147.5  & 10.00 (ref.) & --- \\
G2-E4S1H4                 & 984904.7  & 26.68 & 16.01 \\
G2-H4S5                   & 3287537.4 & 89.06 & 53.44 \\
G2-K4S5                   & 2829374.5 & 76.64 & 45.99 \\
\hline
\end{tabular}
\end{table}

\begin{table}[h!]
\centering
\caption{Approximate molar ratio between phosphorus and each 
G2-peptide conjugate (P:c) determined by $^{31}$P NMR.}
\label{tab:NMR_ratio}
\begin{tabular}{lccc}
\hline
\textbf{Conjugate} & 
\textbf{$n_{\mathrm{conjugate}}$ (\si{\micro\mole})} & 
\textbf{$n_{\mathrm{P}}$ (\si{\micro\mole})} & 
\textbf{P:c ratio} \\
\hline
G2-E4S1H4 & 1.26 & 16.01 & 12.7 \\
G2-H4S5   & 1.20 & 53.44 & 44.5 \\
G2-K4S5   & 1.20 & 45.99 & 38.2 \\
\hline
\end{tabular}
\end{table}

\end{document}

%% file: head.tex
\usepackage{amssymb}
\usepackage{siunitx}
\usepackage{amsmath}
\usepackage[utf8]{inputenc} 

\newcommand*{\mytitle}{Short Peptide Tails Modulate DNA Association and Condensation by PAMAM Dendrimers}

\newcommand*{\pablo}{Pablo M. Blanco}
\newcommand*{\rita}{Rita S. Dias}
\newcommand*{\corina}{Corinna Dannert}
\newcommand*{\sebas}{Sebastian P. Pineda}
\newcommand*{\peter}{Peter Ko\v{s}ovan}

\affiliation[NTNU]{organization={Department of Physics, NTNU - Norwegian University of Science and Technology},
            city={Trondheim},
            postcode={NO-7491}, 
            country={Norway.}}
\affiliation[CUNI]{organization={Department of Physical and Macromolecular Chemistry, Faculty of Science, Charles University},
            addressline={Hlavova 2030}, 
            city={Prague 2},
            postcode={128 40}, 
            country={Czech Republic.}}

\newcommand*{\reffig}[1]{Fig.~\ref{#1}}

\newcommand*{\refeq}[1]{Eq.~\ref{#1}}

\newcommand{\rmol}[0]{$r_{\rm molar}$}
\newcommand{\pKa}[0]{pK$_{\text{a}}$}

\usepackage{tgbonum}
\usepackage{geometry}
\usepackage{parskip}
\usepackage{booktabs}
\usepackage[colorinlistoftodos]{todonotes}
\DeclareSIUnit{\mole}{mol}
\DeclareSIUnit{\molar}{M}
\DeclareSIUnit\angstrom{\text {Å}}
\usepackage{tabularx,booktabs}
\newcolumntype{Y}{>{\centering\arraybackslash}X}

%% file: references.bib
@article{Mendes2017DendrimersTherapy,
    title = {{Dendrimers as nanocarriers for nucleic acid and drug delivery in cancer therapy}},
    year = {2017},
    journal = {Molecules},
    author = {Mendes, Livia Palmerston and Pan, Jiayi and Torchilin, Vladimir P.},
    number = {9},
    month = {9},
    pages = {1401},
    volume = {22},
    publisher = {MDPI AG},
    optdoi = {10.3390/MOLECULES22091401},
    optissn = {14203049},
    pmid = {28832535}
}

@article{Segura2001MaterialsDelivery,
    title = {{Materials for non-viral gene delivery}},
    year = {2001},
    journal = {Annual Review of Materials Science},
    author = {Segura, Tatiana and Shea, Lonnie D.},
    number = {Volume 31, 2001},
    month = {8},
    pages = {25--46},
    volume = {31},
    publisher = {Annual Reviews},
    opturl = {https://www.annualreviews.org/content/journals/10.1146/annurev.matsci.31.1.25},
    optdoi = {10.1146/ANNUREV.MATSCI.31.1.25/CITE/REFWORKS},
    optissn = {00846600}
}

@article{Haensler1993,
    title = {{Polyamidoamine Cascade Polymers Mediate Efficient Transfection of Cells in Culture}},
    year = {1993},
    journal = {Bioconjugate Chemistry},
    author = {Haensler, Jean and Szoka, Francis C.},
    pages = {372--379},
    volume = {4},
    opturl = {https://pubs.acs.org/optdoi/10.1021/bc00023a012.},
    optdoi = {10.1021/bc00023a012}
}

@article{Bielinska1996RegulationDendrimers,
    title = {{Regulation of in vitro gene expression using antisense oligonucleotides or antisense expression plasmids transfected using starburst PAMAM dendrimers}},
    year = {1996},
    journal = {Nucleic Acids Research},
    author = {Bielinska, Anna and Kukowska-Latallo, Jolanta F. and Johnson, Jennifer and Tomalia, Donald A. and Baker, James R.},
    number = {11},
    month = {6},
    pages = {2176--2182},
    volume = {24},
    optdoi = {10.1093/NAR/24.11.2176},
    optissn = {03051048},
    pmid = {8668551}
}

@article{Bielinska1997TheDNA,
    title = {{The interaction of plasmid DNA with polyamidoamine dendrimers: mechanism of complex formation and analysis of alterations induced in nuclease sensitivity and transcriptional activity of the complexed DNA}},
    year = {1997},
    journal = {Biochimica et Biophysica Acta (BBA) - Gene Structure and Expression},
    author = {Bielinska, Anna U. and Kukowska-Latallo, Jolanta F. and Baker, James R.},
    number = {2},
    month = {8},
    pages = {180--190},
    volume = {1353},
    publisher = {Elsevier},
    optdoi = {10.1016/S0167-4781(97)00069-9},
    optissn = {0167-4781},
    pmid = {9294012}
}

@article{Waite2009,
    title = {{PAMAM-RGD Conjugates Enhance siRNA Delivery Through a Multicellular Spheroid Model of Malignant Glioma}},
    year = {2009},
    journal = {Bioconjugate Chemistry},
    author = {Waite, Carolyn L. and Roth, Charles M.},
    number = {10},
    month = {10},
    pages = {1908--1916},
    volume = {20},
    opturl = {https://pubs.acs.org/optdoi/10.1021/bc900228m},
    optdoi = {10.1021/bc900228m},
    optissn = {1043-1802}
}

@article{Tsai2011IntrinsicallyStudy,
    title = {{Intrinsically fluorescent PAMAM dendrimer as gene carrier and nanoprobe for nucleic acids delivery: Bioimaging and transfection study}},
    year = {2011},
    journal = {Biomacromolecules},
    author = {Tsai, Ya Ju and Hu, Chao Chin and Chu, Chih Chien and Imae, Toyoko},
    number = {12},
    month = {12},
    pages = {4283--4290},
    volume = {12},
    publisher = {American Chemical Society},
    opturl = {https://pubs.acs.org/optdoi/full/10.1021/bm201196p},
    optdoi = {10.1021/BM201196P/SUPPL{\_}FILE/BM201196P{\_}SI{\_}001.PDF},
    optissn = {15257797},
    pmid = {22029823}
}

@article{Yoo2000EnhancedDendrimers,
    title = {{Enhanced delivery of antisense oligonucleotides with fluorophore-conjugated PAMAM dendrimers}},
    year = {2000},
    journal = {Nucleic Acids Research},
    author = {Yoo, Hoon and Juliano, R. L.},
    number = {21},
    month = {11},
    pages = {4225--4231},
    volume = {28},
    optdoi = {10.1093/NAR/28.21.4225},
    optissn = {03051048},
    pmid = {11058121}
}

@article{Fu2016DrivingElectrostatic,
    title = {{Driving forces for oppositely charged polyion association in aqueous solutions: enthalpic, entropic, but not electrostatic}},
    year = {2016},
    journal = {Journal of the American Chemical Society},
    author = {Fu, Jingcheng and Schlenoff, Joseph B},
    number = {3},
    pages = {980--990},
    volume = {138},
    publisher = {ACS Publications},
    optissn = {0002-7863}
}

@article{Janaszewska2019CytotoxicityDendrimers,
    title = {{Cytotoxicity of Dendrimers}},
    year = {2019},
    journal = {Biomolecules 2019, Vol. 9, Page 330},
    author = {Janaszewska, Anna and Lazniewska, Joanna and Trzepi{\'{n}}ski, Przemysław and Marcinkowska, Monika and Klajnert-Maculewicz, Barbara},
    number = {8},
    month = {8},
    pages = {330},
    volume = {9},
    publisher = {Multidisciplinary Digital Publishing Institute},
    opturl = {https://www.mdpi.com/2218-273X/9/8/330/htm https://www.mdpi.com/2218-273X/9/8/330},
    optdoi = {10.3390/BIOM9080330},
    optissn = {2218-273X},
    pmid = {31374911}
}

@article{Scott2014EndosomeFunctions,
    title = {{Endosome maturation, transport and functions}},
    year = {2014},
    journal = {Seminars in Cell {\&} Developmental Biology},
    author = {Scott, Cameron C. and Vacca, Fabrizio and Gruenberg, Jean},
    month = {7},
    pages = {2--10},
    volume = {31},
    publisher = {Academic Press},
    optdoi = {10.1016/J.SEMCDB.2014.03.034},
    optissn = {1084-9521},
    pmid = {24709024}
}

@article{Pang2020TheNanoparticles,
    title = {{The Establishment and Application Studies on Precise Lysosome pH Indicator Based on Self-Decomposable Nanoparticles}},
    year = {2020},
    journal = {Nanoscale Research Letters},
    author = {Pang, Cui and Song, Chaojun and Li, Yize and Wang, Qiaofeng and Zhu, Xiaosheng and Wu, Jianwei and Tian, Yi and Fan, Hao and Hu, Jinwei and Li, Chen and Wang, Baolong and Li, Xiaoye and Liu, Wenchao and Fan, Li},
    number = {1},
    month = {7},
    pages = {1--14},
    volume = {15},
    publisher = {Springer},
    opturl = {https://link.springer.com/article/10.1186/s11671-020-03367-0},
    optdoi = {10.1186/S11671-020-03367-0/TABLES/2},
    optissn = {1556276X}
}

@article{Behr1997TheExploit,
    title = {{The Proton Sponge: a Trick to Enter Cells the Viruses Did Not Exploit}},
    year = {1997},
    journal = {CHIMIA},
    author = {Behr, Jean Paul},
    number = {1-2},
    month = {2},
    pages = {34},
    volume = {51},
    publisher = {Swiss Chemical Society},
    opturl = {https://www.chimia.ch/chimia/article/view/2792},
    optdoi = {10.2533/chimia.1997.34},
    optissn = {0009-4293}
}

@article{Won2009MissingComplexes,
    title = {{Missing Pieces in Understanding the Intracellular Trafficking of Polycation/DNA Complexes}},
    year = {2009},
    journal = {Journal of Controlled Release},
    author = {Won, You Yeon and Sharma, Rahul and Konieczny, Stephen F.},
    number = {2},
    month = {10},
    pages = {88},
    volume = {139},
    publisher = {NIH Public Access},
    opturl = {/pmc/articles/PMC3336099/ https://www.ncbi.nlm.nih.gov/pmc/articles/PMC3336099/},
    optdoi = {10.1016/J.JCONREL.2009.06.031},
    optissn = {01683659},
    pmid = {19580830}
}

@article{Needham1990ElasticCholesterol.,
    title = {{Elastic deformation and failure of lipid bilayer membranes containing cholesterol.}},
    year = {1990},
    journal = {Biophysical Journal},
    author = {Needham, David and Nunn, Rashmi S.},
    number = {4},
    pages = {997},
    volume = {58},
    publisher = {The Biophysical Society},
    opturl = {/pmc/articles/PMC1281045/?report=abstract https://www.ncbi.nlm.nih.gov/pmc/articles/PMC1281045/},
    optdoi = {10.1016/S0006-3495(90)82444-9},
    optissn = {00063495},
    pmid = {2249000}
}

@article{Roy2020LysosomalMicrocapsules,
    title = {{Lysosomal Proton Buffering of Poly(ethylenimine) Measured In Situ by Fluorescent pH-Sensor Microcapsules}},
    year = {2020},
    journal = {ACS nano},
    author = {Roy, Sathi and Zhu, Dingcheng and Parak, Wolfgang J. and Feliu, Neus},
    number = {7},
    month = {7},
    pages = {8012--8023},
    volume = {14},
    publisher = {ACS Nano},
    opturl = {https://pubmed.ncbi.nlm.nih.gov/32568521/},
    optdoi = {10.1021/ACSNANO.9B10219},
    optissn = {1936-086X},
    pmid = {32568521}
}

@article{Bieber2002IntracellularComplexes,
    title = {{Intracellular route and transcriptional competence of polyethylenimine–DNA complexes}},
    year = {2002},
    journal = {Journal of Controlled Release},
    author = {Bieber, Thorsten and Meoptissner, Wolfgang and Kostin, Sawa and Niemann, Axel and Elsasser, Hans Peter},
    number = {2-3},
    month = {8},
    pages = {441--454},
    volume = {82},
    publisher = {Elsevier},
    optdoi = {10.1016/S0168-3659(02)00129-3},
    optissn = {0168-3659},
    pmid = {12175756}
}

@article{Vaidyanathan2015QuantitativeMembrane,
    title = {{Quantitative Measurement of Cationic Polymer Vector and Polymer-pDNA Polyplex Intercalation into the Cell Plasma Membrane}},
    year = {2015},
    journal = {ACS Nano},
    author = {Vaidyanathan, Sriram and Anderson, Kevin B. and Merzel, Rachel L. and Jacobovitz, Binyamin and Kaushik, Milan P. and Kelly, Christina N. and Van Dongen, Mallory A. and Dougherty, Casey A. and Orr, Bradford G. and Banaszak Holl, Mark M.},
    number = {6},
    month = {6},
    pages = {6097--6109},
    volume = {9},
    publisher = {American Chemical Society},
    opturl = {https://pubs.acs.org/optdoi/full/10.1021/acsnano.5b01263},
    optdoi = {10.1021/ACSNANO.5B01263/SUPPL{\_}FILE/NN5B01263{\_}SI{\_}001.PDF},
    optissn = {1936086X},
    pmid = {25952271}
}

@article{Rehman2013MechanismLysis,
    title = {{Mechanism of polyplex- and lipoplex-mediated delivery of nucleic acids: Real-time visualization of transient membrane destabilization without endosomal lysis}},
    year = {2013},
    journal = {ACS Nano},
    author = {Rehman, Zia Ur and Hoekstra, Dick and Zuhorn, Inge S.},
    number = {5},
    month = {5},
    pages = {3767--3777},
    volume = {7},
    publisher = {American Chemical Society},
    opturl = {https://pubs.acs.org/optdoi/full/10.1021/nn3049494},
    optdoi = {10.1021/NN3049494/SUPPL{\_}FILE/NN3049494{\_}SI{\_}010.AVI},
    optissn = {19360851},
    pmid = {23597090}
}

@article{Santos2010,
    title = {{Receptor-Mediated Gene Delivery Using PAMAM Dendrimers Conjugated with Peptides Recognized by Mesenchymal Stem Cells}},
    year = {2010},
    journal = {Molecular Pharmaceutics},
    author = {Santos, José L. and Pandita, Deepti and Rodrigues, João and P{\^{e}}go, Ana P. and Granja, Pedro L. and Balian, Gary and Tom{\'{a}}s, Helena},
    number = {3},
    month = {6},
    pages = {763--774},
    volume = {7},
    publisher = {American Chemical Society},
    opturl = {https://pubs.acs.org/optdoi/10.1021/mp9002877},
    optdoi = {10.1021/mp9002877},
    optissn = {1543-8384}
}

@article{Weik2019ESPResSoSystems,
    title = {{ESPResSo 4.0 – an extensible software package for simulating soft matter systems}},
    year = {2019},
    journal = {The European Physical Journal},
    author = {Weik, Florian and Weeber, Rudolf and Szuttor, Kai and Breitsprecher, Konrad and de Graaf, Joost and Kuron, Michael and Landsgesell, Jonas and Menke, Henri and Sean, David and Holm, Christian},
    number = {14},
    month = {3},
    pages = {1789--1816},
    volume = {227}
}

@article{Beyer2024PyMBE:ESPResSo,
    title = {{pyMBE: the Python-based Molecule Builder for ESPResSo}},
    year = {2024},
    journal = {Journal of Chemical Physics},
    author = {Beyer, David and Torres, Paola B. and Pineda, Sebastian P. and Narambuena, Claudio F. and Grad, Jean-Noël and Ko{\v{s}}ovan, Peter and Blanco, Pablo M.},
    number = {2},
    month = {1},
    volume = {161},
    publisher = {American Institute of Physics},
    opturl = {http://arxiv.org/abs/2401.14954 http://dx.optdoi.org/10.1063/5.0216389},
    optdoi = {10.1063/5.0216389},
    arxivId = {2401.14954v3}
}

@article{Sjalander2019EPIC:Infrastructure,
    title = {{EPIC: An Energy-Efficient, High-Performance GPGPU Computing Research Infrastructure}},
    year = {2019},
    author = {Sj{\"{a}}lander, Magnus and Jahre, Magnus and Tufte, Gunnar and Reissmann, Nico},
    month = {12},
    opturl = {https://arxiv.org/abs/1912.05848v5},
    optdoi = {10.48550/arxiv.1912.05848},
    arxivId = {1912.05848}
}

@article{Dannert2023DNACharge,
    title = {{DNA condensation by peptide-conjugated PAMAM dendrimers. influence of peptide charge}},
    year = {2023},
    journal = {ACS omega},
    author = {Dannert, Corinna and Mardal, Ingrid and Lale, Rahmi and Stokke, Bjørn Torger and Dias, Rita S},
    number = {47},
    pages = {44624--44636},
    volume = {8},
    publisher = {ACS Publications},
    optissn = {2470-1343}
}

@article{beyer2025charge,
  title={Charge regulation effects in weak polyelectrolyte complexation},
  author={Beyer, David and Holm, Christian and Wang, Zhen-Gang},
  journal={The Journal of Physical Chemistry Letters},
  volume={16},
  number={32},
  pages={8245--8251},
  year={2025},
  publisher={ACS Publications}
}

@Article{Keri2017,
author ="Kéri, Mónika and Nagy, Zoltán and Novák, Levente and Szarvas, Edit and Balogh, Lajos P. and Bányai, István",
title  ="Beware of phosphate: evidence of specific dendrimer–phosphate interactions",
journal  ="Physical Chemistry Chemical Physics",
year  ="2017",
volume  ="19",
issue  ="18",
pages  ="11540-11548"
}

@article{Cakara2003MicroscopicTitrations,
    title = {{Microscopic protonation equilibria of poly(amidoamine) dendrimers from macroscopic titrations}},
    year = {2003},
    journal = {Macromolecules},
    author = {Cakara, Dusko and Kleimann, Jörg and Borkovec, Michal},
    number = {11},
    month = {6},
    pages = {4201--4207},
    volume = {36},
    publisher = { American Chemical Society },
    opturl = {https://pubs.acs.org/optdoi/full/10.1021/ma0300241},
    optdoi = {10.1021/MA0300241/ASSET/IMAGES/LARGE/MA0300241F00007.JPEG},
    optissn = {00249297}
}

@article{Pineda2024ChargePolyelectrolytes,
    title = {{Charge Regulation Triggers Condensation of Short Oligopeptides to Polyelectrolytes}},
    year = {2024},
    journal = {JACS Au},
    author = {Pineda, Sebastian P. and Staňo, Roman and Murmiliuk, Anastasiia and Blanco, Pablo M. and Montes, Patricia and To{\v{s}}ner, Zdeněk and Groborz, Ondřej and P{\'{a}}nek, Jiří and Hrub{\'{y}}, Martin and {\v{S}}t{\v{e}}p{\'{a}}nek, Miroslav and Ko{\v{s}}ovan, Peter},
    number = {5},
    month = {5},
    pages = {1775--1785},
    volume = {4},
    publisher = {American Chemical Society},
    opturl = {/optdoi/pdf/10.1021/jacsau.3c00668},
    optdoi = {10.1021/JACSAU.3C00668/ASSET/IMAGES/LARGE/AU3C00668{\_}0005.JPEG},
    optissn = {26913704}
}

@article{Pettersen2004UCSFAnalysis,
    title = {{UCSF Chimera - A visualization system for exploratory research and analysis}},
    year = {2004},
    journal = {Journal of Computational Chemistry},
    author = {Pettersen, Eric F. and Goddard, Thomas D. and Huang, Conrad C. and Couch, Gregory S. and Greenblatt, Daniel M. and Meng, Elaine C. and Ferrin, Thomas E.},
    number = {13},
    month = {10},
    pages = {1605--1612},
    volume = {25},
    publisher = {J Comput Chem},
    opturl = {https://pubmed.ncbi.nlm.nih.gov/15264254/},
    optdoi = {10.1002/JCC.20084,},
    optissn = {01928651},
    pmid = {15264254}
}

@article{Lunkad2021QuantitativeOligopeptides,
    title = {{Quantitative prediction of charge regulation in oligopeptides}},
    year = {2021},
    journal = {Molecular Systems Design {\&} Engineering},
    author = {Lunkad, Raju and Murmiliuk, Anastasiia and Hebbeker, Pascal and Boubl{\'{i}}k, Milan and To{\v{s}}ner, Zdeněk and {\v{S}}t{\v{e}}p{\'{a}}nek, Miroslav and Ko{\v{s}}ovan, Peter},
    number = {2},
    month = {2},
    pages = {122--131},
    volume = {6},
    publisher = {The Royal Society of Chemistry},
    opturl = {https://pubs.rsc.org/en/content/articlehtml/2021/me/d0me00147c https://pubs.rsc.org/en/content/articlelanding/2021/me/d0me00147c},
    optdoi = {10.1039/D0ME00147C},
    optissn = {2058-9689}
}

@article{Reed1992MontePolyelectrolytes,
    title = {{Monte Carlo study of titration of linear polyelectrolytes}},
    year = {1992},
    journal = {The Journal of chemical physics},
    author = {Reed, Christopher E and Reed, Wayne F},
    number = {2},
    pages = {1609--1620},
    volume = {96},
    publisher = {American Institute of Physics},
    optissn = {0021-9606}
}

@article{Janke2002StatisticalEstimation,
    title = {{Statistical analysis of simulations: Data correlations and error estimation}},
    year = {2002},
    journal = {Quantum simulations of complex many-body systems: from theory to algorithms},
    author = {Janke, Wolfhard},
    pages = {423--445},
    volume = {10},
    publisher = {Citeseer}
}

@article{Souaille2001ExtensionCalculations,
    title = {{Extension to the weighted histogram analysis method: combining umbrella sampling with free energy calculations}},
    year = {2001},
    journal = {Computer physics communications},
    author = {Souaille, Marc and Roux, Benoıt},
    number = {1},
    pages = {40--57},
    volume = {135},
    publisher = {Elsevier},
    optissn = {0010-4655}
}

@article{dia03:8150,
    title = {{Modeling of DNA compaction by polycations}},
    year = {2003},
    journal = {J. Chem. Phys.},
    author = {Dias, Rita S. and Pais, Alberto A. C. C. and Lindman, Björn and Miguel, Maria G.},
    pages = {8150--8157},
    volume = {119}
}

@article{ras98:381,
    title = {{Precipitation of DNA by polyamines: A polyelectrolyte behavior}},
    year = {1998},
    journal = {Biophys. J.},
    author = {Raspaud, Eric and Cruz, Monica O. and Shikorav, Jean-Louis and Livolant, Francoise},
    pages = {381--393},
    volume = {74}
}

@article{Takahashi1997DiscretePolyamines,
    title = {{Discrete coil-globule transition of single duplex DNAs induced by polyamines}},
    year = {1997},
    journal = {Journal of Physical Chemistry B},
    author = {Takahashi, Masazumi and Yoshikawa, Kenichi and Vasilevskaya, Valentina V. and Khokhlov, Alexei R.},
    number = {45},
    month = {11},
    pages = {9396--9401},
    volume = {101},
    publisher = {American Chemical Society},
    opturl = {https://pubs.acs.org/optdoi/full/10.1021/jp9716391},
    optdoi = {10.1021/JP9716391/ASSET/IMAGES/MEDIUM/JP9716391E00006.GIF},
    optissn = {15206106}
}

@article{Ullner1994ConformationalArguments,
    title = {{Conformational properties and apparent dissociation constants of titrating polyelectrolytes: Monte Carlo simulation and scaling arguments}},
    year = {1994},
    journal = {The Journal of Chemical Physics},
    author = {Ullner, Magnus and J{\"{o}}nsson, Bo and Widmark, Per Olof},
    number = {4},
    month = {2},
    pages = {3365--3366},
    volume = {100},
    publisher = {AIP Publishing},
    opturl = {/aip/jcp/article/100/4/3365/112623/Conformational-properties-and-apparent},
    optdoi = {10.1063/1.466378},
    optissn = {0021-9606}
}

@article{Blanco2023UnusualPolyelectrolytes,
    title = {{Unusual aspects of charge regulation in flexible weak polyelectrolytes}},
    year = {2023},
    journal = {Polymers},
    author = {Blanco, Pablo M and Narambuena, Claudio F and Madurga, Sergio and Mas, Francesc and Garc{\'{e}}s, Josep L},
    number = {12},
    pages = {2680},
    volume = {15},
    publisher = {Multidisciplinary Digital Publishing Institute},
    optissn = {2073-4360}
}

@article{Landsgesell2019SimulationsGels,
    title = {{Simulations of ionization equilibria in weak polyelectrolyte solutions and gels}},
    year = {2019},
    journal = {Soft Matter},
    author = {Landsgesell, Jonas and Nov{\'{a}}, Lucie and Rud, Oleg and Uhl{\'{i}}k, Filip and Sean, David and Hebbeker, Pascal and Holm, Christian and Ko{\v{s}}ovan, Peter},
    number = {6},
    month = {2},
    pages = {1155--1185},
    volume = {15},
    publisher = {Royal Society of Chemistry},
    opturl = {https://pubs.rsc.org/en/content/articlehtml/2019/sm/c8sm02085j https://pubs.rsc.org/en/content/articlelanding/2019/sm/c8sm02085j},
    optdoi = {10.1039/C8SM02085J},
    optissn = {17446848},
    pmid = {30706070}
}

@article{Stornes2017MonteConcentration,
    title = {{Monte Carlo Simulations of Complexation between Weak Polyelectrolytes and a Charged Nanoparticle. Influence of Polyelectrolyte Chain Length and Concentration}},
    year = {2017},
    journal = {Macromolecules},
    author = {Stornes, Morten and Linse, Per and Dias, Rita S.},
    number = {15},
    pages = {5978--5988},
    volume = {50},
    address = {NTNU Norwegian Univ Sci {\&} Technol, Dept Phys, NO-7491 Trondheim, Norway Lund Univ, Div Phys Chem, Ctr Chem {\&} Chem Engn, S-22100 Lund, Sweden},
    optdoi = {10.1021/acs.macromol.7b00844},
    optissn = {0024-9297},
    language = {English}
}

@article{Ulrich2005ComplexationInfluences,
    title = {{Complexation of a weak polyelectrolyte with a charged nanoparticle. Solution properties and polyelectrolyte stiffness influences}},
    year = {2005},
    journal = {Macromolecules},
    author = {Ulrich, Serge and Laguecir, Abohachem and Stoll, Serge},
    number = {21},
    month = {10},
    pages = {8939--8949},
    volume = {38},
    publisher = { American Chemical Society },
    opturl = {/optdoi/pdf/10.1021/ma051142m?ref=article_openPDF},
    optdoi = {10.1021/MA051142M/ASSET/IMAGES/LARGE/MA051142MF00010.JPEG},
    optissn = {00249297}
}

@article{Rathee2018WeakCharging,
    title = {{Weak polyelectrolyte complexation driven by associative charging}},
    year = {2018},
    journal = {Journal of Chemical Physics},
    author = {Rathee, Vikramjit S. and Zervoudakis, Aristotle J. and Sidky, Hythem and Sikora, Benjamin J. and Whitmer, Jonathan K.},
    number = {11},
    month = {3},
    pages = {114901},
    volume = {148},
    publisher = {American Institute of Physics Inc.},
    opturl = {/aip/jcp/article/148/11/114901/195864/Weak-polyelectrolyte-complexation-driven-by},
    optdoi = {10.1063/1.5017941/13335561/114901{\_}1{\_}ACCEPTED{\_}MANUSCRIPT.PDF},
    optissn = {00219606},
    pmid = {29566508}
}

@article{NarayananNair2017ComplexationPolyampholytes,
    title = {{Complexation Behavior of Polyelectrolytes and Polyampholytes}},
    year = {2017},
    journal = {Journal of Physical Chemistry B},
    author = {Narayanan Nair, Arun Kumar and Martinez Jimenez, Arturo and Sun, Shuyu},
    number = {33},
    month = {8},
    pages = {7987--7998},
    volume = {121},
    publisher = {American Chemical Society},
    opturl = {/optdoi/pdf/10.1021/acs.jpcb.7b04582?ref=article_openPDF},
    optdoi = {10.1021/ACS.JPCB.7B04582/ASSET/IMAGES/LARGE/JP-2017-04582N{\_}0010.JPEG},
    optissn = {15205207},
    pmid = {28742350}
}

@article{Ainalem2009CondensingMorphology,
    title = {{Condensing DNA with poly(amido amine) dendrimers of different generations: means of controlling aggregate morphology}},
    year = {2009},
    journal = {Soft Matter},
    author = {Ainalem, Marie Louise and Carnerup, Anna Margareta and Janiak, John and Alfredsson, Vivekae and Nylander, Tommy and Schill{\'{e}}n, Karin},
    number = {11},
    month = {5},
    pages = {2310--2320},
    volume = {5},
    publisher = {Royal Society of Chemistry},
    opturl = {https://pubs.rsc.org/en/content/articlehtml/2009/sm/b821629k https://pubs.rsc.org/en/content/articlelanding/2009/sm/b821629k},
    isbn = {10/7/202410:1},
    optdoi = {10.1039/B821629K},
    optissn = {1744683X}
}

@article{Fant2008DNATranscription,
    title = {{DNA condensation by PAMAM dendrimers: Self-assembly characteristics and effect on transcription}},
    year = {2008},
    journal = {Biochemistry},
    author = {Fant, Kristina and Esbj{\"{o}}rner, Elin K. and Lincoln, Per and Nord{\'{e}}n, Bengt},
    number = {6},
    month = {2},
    pages = {1732--1740},
    volume = {47},
    optdoi = {10.1021/BI7017199},
    optissn = {00062960},
    pmid = {18189415}
}

@article{Bae2021ApoptinTherapy,
    title = {{Apoptin gene delivery by a PAMAM dendrimer modified with a nuclear localization signal peptide as a gene carrier for brain cancer therapy}},
    year = {2021},
    journal = {Korean Journal of Physiology and Pharmacology},
    author = {Bae, Yoonhee and Lee, Jell and Kho, Changwon and Choi, Joon Sig and Han, Jin},
    number = {5},
    month = {9},
    pages = {467--478},
    volume = {25},
    publisher = {The Korean Physiological Society; The Korean Society of Pharmacology},
    opturl = {http://dx.optdoi.org/10.4196/kjpp.2021.25.5.467 http://www.e-sciencecentral.org/articles/?scid=1147417},
    optdoi = {10.4196/kjpp.2021.25.5.467},
    optissn = {1226-4512}
}

@article{Ebrahimian2022DevelopmentStudies,
    title = {{Development of targeted gene delivery system based on liposome and PAMAM dendrimer functionalized with hyaluronic acid and TAT peptide: In vitro and in vivo studies}},
    year = {2022},
    journal = {Biotechnology Progress},
    author = {Ebrahimian, Mahboubeh and Hashemi, Maryam and Farzadnia, Mahdi and Zarei-Ghanavati, Siamak and Malaekeh-Nikouei, Bizhan},
    number = {5},
    month = {9},
    pages = {e3278},
    volume = {38},
    publisher = {American Chemical Society (ACS)},
    opturl = {https://optdoi.org/10.1002/btpr.3278},
    optdoi = {https://optdoi.org/10.1002/btpr.3278},
    optissn = {8756-7938}
}

@article{Urbiola2018NovelReceptors,
    title = {{Novel PAMAM-PEG-Peptide Conjugates for siRNA Delivery Targeted to the Transferrin and Epidermal Growth Factor Receptors}},
    year = {2018},
    journal = {Journal of Personalized Medicine 2018, Vol. 8, Page 4},
    author = {Urbiola, Koldo and Blanco-Fern{\'{a}}ndez, Laura and Ogris, Manfred and R{\"{o}}dl, Wolfgang and Wagner, Ernst and de Ilarduya, Conchita Tros},
    number = {1},
    month = {1},
    pages = {4},
    volume = {8},
    publisher = {Multidisciplinary Digital Publishing Institute},
    opturl = {https://www.mdpi.com/2075-4426/8/1/4/htm https://www.mdpi.com/2075-4426/8/1/4},
    optdoi = {10.3390/JPM8010004},
    optissn = {2075-4426}
}

@article{gul84:2221,
    title = {{Electric double layer forces. A Monte Carlo study}},
    year = {1984},
    journal = {Journal of Chemical Physics},
    author = {Gulbrandt, Lars and J{\"{o}}nsson, Bo and Wennerstr{\"{o}}m, Håkan and Linse, Per},
    pages = {2221--2226},
    volume = {88}
}

@article{kha99:121,
    title = {{Electrostatic correlations fold DNA}},
    year = {1999},
    journal = {Biopolymers},
    author = {Khan, Malek O. and J{\"{o}}nsson, Bo},
    pages = {121--125},
    volume = {49}
}

@article{Nozaki196784Behavior,
    title = {{[84] Examination of titration behavior}},
    year = {1967},
    journal = {Methods in Enzymology},
    author = {Nozaki, Yasuhiko and Tanford, Charles},
    number = {C},
    month = {1},
    pages = {715--734},
    volume = {11},
    publisher = {Academic Press},
    opturl = {https://www.sciencedirect.com/science/article/abs/pii/S0076687967110884?via%3Dihub},
    optdoi = {10.1016/S0076-6879(67)11088-4},
    optissn = {0076-6879}
}

@article{Khan1999AnomalousTheory,
    title = {{Anomalous salt effects on DNA conformation: Experiment and theory}},
    year = {1999},
    journal = {Macromolecules},
    author = {Khan, Malek O. and Mel'nikov, Sergey M. and J{\"{o}}nsson, Bo},
    number = {26},
    month = {12},
    pages = {8836--8840},
    volume = {32},
    publisher = {ACS},
    opturl = {/optdoi/pdf/10.1021/ma9905627?ref=article_openPDF},
    optdoi = {10.1021/MA9905627/ASSET/IMAGES/LARGE/MA9905627F00006.JPEG},
    optissn = {00249297}
}

@article{Mantelli2011,
    author = {Mantelli, S. and Muller, P. and Harlepp, S. and Maaloum, M.},
    title = {Conformational analysis and estimation of the persistence length of DNA using atomic force microscopy in solution},
    journal = {Soft Matter},
    volume = {7},
    number = {7},
    pages = {3412-3416},
    year = {2011}
}

@article{Bus2018,
    author = {Bus, Tanja and Traeger, Anja and Schubert, Ulrich S.},
    title = {The great escape: how cationic polyplexes overcome the endosomal barrier},
    journal = {Journal of Materials Chemistry B},
    volume = {6},
    number = {43},
    pages = {6904-6918},
    year = {2018},
    month = {11}
}

@article{PAMAM_Subm,
    title = {{Electrostatic interactions drive the accumulation of phosphate ions into cationic dendrimers}},
    year = {2026},
    author = {Blanco, Pablo M. and Balaji, Ganesh and Dannert, Corina and Pineda, Sebastian P. and Rey-Castro, Carlos and Garc\'es, Josep Llu\'is and Ko{\v{s}}ovan, Peder and Dias, Rita S.},
    journal = {Submitted},}

@Article{beyer25a,
author = {Beyer, David and Blanco, Pablo M. and Landsgesell, Jonas and Košovan, Peter and Holm, Christian},
title = {How To Correct Erroneous Symmetry-Breaking in Coarse-Grained Constant-pH Simulations},
journal = {Journal of Chemical Theory and Computation},
volume = {21},
number = {3},
pages = {1396-1404},
year = {2025}
}

@InCollection{labbez07b,
  author    = {Labbez, Christophe and J{\"o}nsson, Bo},
  title     = {A New {M}onte {C}arlo Method for the Titration of Molecules and Minerals},
  booktitle = {Applied Parallel Computing. State of the Art in Scientific Computing},
  year      = {2007},
  editor    = {K{\aa}gstr{\"o}m, Bo and Elmroth, Erik and Dongarra, Jack and Wa{\'s}niewski, Jerzy},
  volume    = {4699},
  series    = {Lecture Notes in Computer Science},
  publisher = {Springer Berlin Heidelberg},
  pages     = {66--72}
}
